\documentclass[12pt,letterpaper]{article}
\usepackage{jcappub}
\newcommand{\Author}[2][a]{\author[#1]{#2}}
\newcommand{\Affiliation}[2][a]{\affiliation[#1]{#2}}
\newcommand{\Email}{\emailAdd}
\newenvironment{Abstract}{\startabstract}{}
\def\startabstract#1\end{\abstract{#1}\end}
\newcommand{\lsim}{\mathrel{\mathop{\kern 0pt \rlap
      {\raise.2ex\hbox{$<$}}}\lower.9ex\hbox{\kern-.190em $ \sim$}}}
\newcommand{\gsim}{\mathrel{\mathop{\kern 0pt
      \rlap{\raise.2ex\hbox{$>$}}}\lower.9ex\hbox{\kern-.190em $\sim$}}}

\usepackage{amsmath}
\usepackage{amssymb}
\usepackage{mathrsfs}
\usepackage{amsbsy}
\usepackage{amsfonts}
\usepackage{latexsym}
\usepackage{graphicx}
\usepackage{epstopdf}
\usepackage{color}
\usepackage{bm}
\usepackage[normalem]{ulem}
\usepackage{cancel}
\usepackage{afterpage}
\usepackage{float}

\newenvironment{smalltext}{\color[RGB]{0,110,0}\small}{}
\newcommand{\nosmalltext}{\renewenvironment{smalltext}{\gobble\bgroup}{\egroup}}

\newenvironment{yellow}{\bgroup\color{yellow}}{\egroup}

\newcommand{\scrH}{{\cal H}}

\newcommand{\scrG}{{\cal G}}

\newcommand{\scrP}{{\cal P}}

\newcommand{\vv}{{\bf v}}
\newcommand{\VV}{{\bf V}}
\newcommand{\vmin}{v_{\rm min}}
\newcommand{\gal}{{\rm gal}}
\newcommand{\lab}{{\rm lab}}
\newcommand{\thr}{{\rm thr}}
\newcommand{\esc}{{\rm esc}}

\newcommand{\uu}{{\bf u}}

\newcommand{\sun}{\odot}
\newcommand{\earth}{\oplus}
\newcommand{\vsun}{v_\sun}
\newcommand{\vearth}{v_\earth}
\newcommand{\vvsun}{{\bf v}_\sun}
\newcommand{\vvearth}{{\bf v}_\earth}

\allowdisplaybreaks 

\begin{document}

\title{Halo--independent determination of the unmodulated WIMP
  signal in DAMA: the isotropic case.}

\Author[a]{Paolo Gondolo}
\Email{paolo.gondolo@utah.edu}
\Affiliation[a]{Department of Physics, University of Utah, 115 South 1400
  East \#201, Salt Lake City, Utah 84112-0830}
\Author[b]{Stefano Scopel}
\Email{scopel@sogang.ac.kr}
\Affiliation[b]{Department of Physics, Sogang University, Seoul 121-742, South Korea}

\begin{Abstract}
  We present a halo-independent determination of the unmodulated
  signal corresponding to the DAMA modulation if interpreted as due to
  dark matter weakly interacting massive particles (WIMPs). First we
  show how a modulated signal gives information on the WIMP velocity
  distribution function in the Galactic rest frame from which the
  unmodulated signal descends. Then we describe a mathematically-sound
  profile likelihood analysis in which the likelihood is profiled over
  a continuum of nuisance parameters (namely, the WIMP velocity
  distribution). As a first application of the method, which is very
  general and valid for any class of velocity distributions, we
  restrict the analysis to velocity distributions that are isotropic
  in the Galactic frame. In this way we obtain halo-independent
  maximum-likelihood estimates and confidence intervals for the DAMA
  unmodulated signal. We find that the estimated unmodulated signal is
  in line with expectations for a WIMP-induced modulation and is
  compatible with the DAMA background+signal rate. Specifically, for the
  isotropic case we find that the modulated amplitude ranges between a
  few percent and about 25\% of the unmodulated amplitude, depending
  on the WIMP mass.
\end{Abstract}

\maketitle

\section{Introduction}

Discovering the nature of Dark Matter (DM) is one of the most important
endeavors in today's particle physics and cosmology. Weakly
Interacting Massive Particles (WIMPs), the most popular and natural DM
candidates, provide the correct thermal relic density in the early
Universe when their cross section with Standard Model particles is at
the level of or not much smaller than weak interaction cross sections. Many
experiments are currently trying to exploit this fact to detect the
WIMPs supposed to form the dark halo of our Galaxy through
their scattering off atomic nuclei in low-background laboratory detectors. An expected feature of halo WIMP scattering is an annual modulation of the scattering rate due to the revolution of the Earth around the Sun \cite{Drukier:1986tm}. 
The importance of such yearly modulation for WIMP
direct detection rests on the fact that, in absence of a sensitivity to the
direction of the incoming particles, it is the only known signature
that allows to distinguish a WIMP signal from the background due to
radioactive contamination, since the latter may have an energy spectrum
indistinguishable from that predicted for WIMPs.

An annual modulation with WIMP characteristics has been claimed for many years by the DAMA experiment \cite{dama}. The low--energy event rate in the DAMA sodium iodide scintillators is well represented by a signal of the form 
\begin{align}
S(t)=S_0+S_m\cos[\omega(t-t_0)],
\label{eq:Smodulation}
\end{align}
with $\omega=2\pi/T$, $T=1~{\rm yr}$, and $t_0\simeq$
2$^{\rm nd}$ of June, as expected for a nonrotating dark halo of WIMPs.
The statistical significance of the DAMA modulation signal
exceeds 9 standard deviations and the effect has been recorded through
14 yearly periods.  For a typical Maxwellian distribution of WIMP
velocities in the Galactic rest frame with rms velocity below 300 km/s, the
modulated component $S_{m}$ of the signal is predicted to be less that
10\% of the unmodulated component $S_0$, in all of the energy bins of the
DAMA detected spectrum. 

The DAMA claim has prompted a world--wide effort by other direct
detection experiments to confirm or disprove the signal
\cite{lux,xenon100,xenon10,kims,cdms_ge,cdms_lite,super_cdms,simple,coupp,picasso,pico2l,pico60,xmass}.
The experiments that have by now reached a background level low enough
to be sensitive to the DAMA modulation use targets different from
sodium iodide. As a consequence, while for standard hypotheses on the
WIMP--nucleon cross section and WIMP velocity distribution (i.e.,
spin--dependent or spin--independent interactions with a truncated
Maxwellian distribution) the DAMA signal appears to be in strong
tension with constraints from other detectors, when such assumptions
are relaxed several WIMP models have been shown to exist for which the
yearly modulation effect measured by DAMA can still be reconciled with
the non observation of a dark matter signal in other experiments
\cite{1405.0364,1502.07682,1505.01926,1701.02215}. This shows that
there is still a clear need to assess the compatibility of the DAMA
excess with other detectors in a model--independent way.

Eliminating the dependence on astrophysics is the underlying goal of the halo--independent approach. Its first formulation~\cite{1011.1915} was based on the observation that the elastic spin-independent scattering rate of WIMPs in a detector depends on the velocity distribution only through a single velocity integral $\tilde{\eta}(\vmin)$, the same for all experiments,
\begin{align}
\tilde{\eta}(\vmin) =  \frac{\rho_{\chi}}{m_{\chi}} \, \sigma_{\chi N} \int_{|\vv|>\vmin}\frac{f_\lab(\vv)}{|\vv|}\, d^3 v .
\label{eq:eta_tilde}
\end{align}
Here $m_\chi$ is the WIMP mass, $\sigma_{\chi N}$ is the WIMP-nucleon
cross section, $\rho_\chi$ is the local WIMP mass density and
$f_{\lab}(\vv)$ is the WIMP velocity distribution in the frame of the
detector (the laboratory). The method of~\cite{1011.1915} has been
applied to the comparison of experiments in
\cite{1107.0717,1107.0741,1011.1915,1111.0292,1112.1627,1205.0134,1304.6066,1405.1420,1502.03342,1504.03333}.
It has been generalized to arbitrary WIMP--nucleon interactions and
any direct detection experiment (with arbitrary efficiency and energy
resolution) in~\cite{1202.6359} by defining weighted averages of
$\tilde{\eta}(\vmin)$ over the range(s) of velocities measured in an
experiment. Applications of the latter method to the comparison of
experiments for various WIMP--nucleon interactions can be found
in~\cite{1202.6359,1304.6183,1311.4247,1401.4508,1404.7484,1405.0364,1405.5582,1502.07682,1505.01926,1507.03902,1607.02445,1703.06892}. Maximum-likelihood
methods to determine the velocity integral and particle physics
parameters have been used
in~\cite{1403.4606,1403.6830,1405.1420,1409.5446,1504.03333,1507.03902,1607.02445,1607.04418,1703.06892}
and statistical methods to assess the compatibility of experiments
have been considered in~\cite{1409.5446,1410.6160,1607.02445}.
Alternative methods to place halo--independent bounds on particle
physics parameters have been put forward
in~\cite{Drees:2007hr,0803.4477,1011.1910,1003.5283,1103.5145,1207.2039,1303.6868,1310.7039,1312.1852, 1502.04224,1506.03386,1609.08630,IbarraRappeltMIAPP}.


A weak point of the halo-independent methods above is the way they compare modulated and unmodulated rates. Some authors define two separate velocity integrals $\tilde{\eta}_0(\vmin)$ and $\tilde{\eta}_m(\vmin)$, one for the unmodulated part $S_0$ and one for the modulated part $S_m$ in Eq.~(\ref{eq:Smodulation}), and then proceed to impose either the simple inequality $\tilde{\eta}_m(\vmin) < \tilde{\eta}_0(\vmin)$ \cite{1202.6359,1304.6183,1311.4247,1401.4508,1404.7484,1405.5582,1502.07682,1505.01926,1507.03902,1607.02445} or  more sophisticated inequalities valid for smooth distributions~\cite{1112.1627,1205.0134}. Comparing two separate velocity integrals with proper statistical significance is not straightforward. Other authors replace the modulation amplitude $S_m$, which is a coefficient in a Fourier time-series, with half the difference between the maximum and minimum signal during a year, i.e., replace $\tilde{\eta}_m(\vmin)$ with $ \tilde{\eta}_{1/2}(\vmin) = [\tilde{\eta}_0(t_0)+\tilde{\eta}_0(t_0+\tfrac{1}{2}T)]/2$ \cite{1111.0292,1304.6066,1405.0364}. This replacement is inaccurate in a halo-independent approach, where one must include velocity distributions for which the modulation is not sinusoidal near the threshold region in which the DAMA signal is present (in this case, the theoretical values of $\tilde{\eta}_{1/2}$ and $S_m$ may be very different, and without a control on the sinusoidal character of the modulation, it would be inappropriate to compare the theoretical $\tilde{\eta}_{1/2}$ with the measured $S_m$).

The main goal of this paper is to show that it is possible to obtain
  information on the unmodulated signal $S_0$ from a measurement of the modulation amplitudes $S_m$ without specifying
  the WIMP velocity distribution (and without assuming two separate velocity integrals or approximating the Fourier coefficient with a difference). For this purpose, (a) we transfer the modulation from being a property of the velocity distribution to being a property of the detector, indeed of the relative motion of the detector with respect to the rest frame of the WIMP population, and (b) we profile the likelihood at fixed $S_0$ over all WIMP velocity distributions (a continuum of nuisance parameters) using rigorous mathematical methods based on the theory of linear optimization in spaces of functions that have the dimension of the continuum. 
  
As a first
application of the method, which is very general and valid for any elastic or inelastic cross section and any
class of velocity distributions, we estimate the DAMA unmodulated signal starting from data on the modulation amplitudes under some simplifying assumptions that have allowed us to explore and understand the difficulties and merits of the method. We restrict the analysis to velocity
distributions that are isotropic in the Galactic frame. We apply our method to the DAMA data for the case of WIMP--nucleus
elastic scattering and for WIMP masses $m_{\chi}<$ 15 GeV, for which
only sodium targets contribute to the expected signal.\footnote{At
  higher WIMP masses the expected rate in DAMA also takes
  contributions from scattering off iodine. This part of the signal is
  in principle constrained in a model--independent way by
  KIMS~\cite{kims}, which uses a CsI detector, and COUPP~\cite{coupp},
  which uses a CF$_3$I detector, since such detectors employ the same
  nuclear target (iodine).} In this way we will obtain quantitative
confidence intervals for the unmodulated components of the WIMP signal
in each energy bin, for the first time disentangling the unmodulated
signal from the background in a halo--independent way.  The $S_0$
confidence intervals we find are valid for any WIMP--nucleus
interaction in which the ${}^{23}$Na cross section
varies negligibly in the 2--4 keVee energy range where the DAMA
modulation is present. This includes the standard elastic spin--independent
and spin--dependent interactions, with arbitrary ratios of the
proton--WIMP and neutron--WIMP coupling constants, but does not
include inelastic scattering or effective WIMP--nucleon operators that show an explicit and fast
dependence on the WIMP--nucleus relative velocity and/or on the
exchanged momentum (like some of those in~\cite{haxton1}).   Since the
properties discussed in the present paper pertain exclusively to
sodium targets in DAMA, while other detectors that constrain DAMA use
different target nuclei, we will not discuss the latter any further.

The plan of this paper is as follows. In Section~\ref{sec:galactic} we show how to
compare the modulated and the unmodulated rates directly in terms of a
single velocity distribution function, namely the velocity distribution function in the Galactic rest frame. 
In Section~\ref{sec:extreme} we present and discuss our method to compute the profile 
likelihood of the unmodulated signal by using linear optimization theory in the continuum to profile the likelihood over the whole velocity distribution (a continuum of nuisance parameters). Section~\ref{sec:analysis} is devoted to our numerical analysis of the DAMA data for velocity distributions that are isotropic in the Galactic
frame. Figs.~\ref{fig:mchi_5_e_s0}, \ref{fig:mchi_10_e_s0} and
\ref{fig:mchi_15_e_s0} and Table \ref{tab:S} of this Section contain our main quantitative results. Finally, the  Appendix contains details of the calculation of the modulated and unmodulated Galactic  response functions for isotropic velocity distributions. 

\section{Rates in terms of the Galactic velocity distribution}
\label{sec:galactic}

Let $S_{i}(t)$ be the expected signal in a dark matter detector, where $t$ is
time, and the index $i$, which may be continuous, 
specifies the quantity measured in the experiment, for example detected recoil energy or energy bin or
number of photoelectrons. Let $f_\lab(\vv,t)$ be the WIMP velocity
distribution function in the frame of the detector, normalized to 
\begin{align}
\int f_\lab(\vv,t) \, d^3v  = 1 .
\end{align}
The signal $S_i(t)$ depends on $f_\lab(\vv,t)$ according to the general formula
\cite{del_nobile_generalized}:
\begin{align}
S_{i}(t) = \int \scrH_{i}(\vv) \, f_\lab(\vv,t) \, d^3v ,
\label{eq:Spt}
\end{align}
where $\scrH_{i}(\vv)$, called the response function, equals the value the signal would have if all the WIMPs had the same velocity $\vv$. Formula (\ref{eq:Spt}) can be understood for example by writing the scattering rate per unit target mass $dR_{\chi T}/dE_R$ of a WIMP $\chi$ off an isotope $T$ in the target, differential in the nucleus recoil energy $E_R$, as a product of the differential cross section $d\sigma_{\chi T}/dE_R$ and the WIMP flux $n_\chi v f_\lab(\vv,t) d^3v$ (where $n_\chi=\rho_\chi/m_\chi$ is the $\chi$ number density), the whole quantity divided by the mass $m_T$ of the target isotope,
\begin{align}
\frac{dR_{\chi T}}{dE_R} = \frac{1}{m_T} \int \frac{d\sigma_{\chi T}}{dE_R} \, \frac{\rho_\chi}{m_\chi} \, v \, f_\lab(\vv,t) \, d^3v .
\label{eq:scattrate}
\end{align}
Here it is understood that the differential cross section $d\sigma_{\chi T}/dE_R$ is  nonzero only in the kinematically allowed region (e.g., for elastic scattering, only for $E_R\le E_R^{\rm max}(v)$ given in equation (\ref{eq:ERmax}) below). Furthermore, if $\scrP_T(E,E_R)$ indicates the probability of actually observing an event with observed energy $E$ when a WIMP has scattered off an isotope $T$ in the detector target with recoil energy $E_R$, the expected observed event rate per unit target mass $dR/dE$ is given by the convolution
\begin{align}
\frac{dR}{dE} = \sum_T \, C_T \! \int \scrP_T(E,E_R) \, \frac{dR_{\chi T}}{dE_R}  \, dE_R .
\label{eq:eventrateconvolution}
\end{align}
Here $C_T $ is the mass fraction of isotope $T$ in the target. Inserting the scattering rate in equation (\ref{eq:scattrate}) into the latter expression, and exchanging the order of the integrations over $E_R$ and $\vv$, leads to the following expression of the response function $\scrH_E(v)$, where $i=E$, for the differential event rate $dR/dE$,
\begin{align}
\scrH_{E}(v) = v \, \frac{\rho_\chi}{m_\chi} \int dE_R \,  \sum_T \scrP_T(E,E_R) \, \frac{C_T}{m_T} \, \frac{d\sigma_{\chi T}}{dE_R}. 
\end{align}
The response function depends on the particle physics model for the interaction of the WIMP with the target and includes the probability that a WIMP scattering in the detector is actually observed. The nonzero values of the response function $\scrH_i(\vv)$ also indicate the WIMP velocities $\vv$ to which the observed signal $S_i(t)$
is sensitive. General expressions of the response functions $\scrH_i(\vv)$ for experiments counting number of events in observed energy bins can be found in \cite{del_nobile_generalized}.

One commonly assumes that the response function $ \scrH_{i}(\vv) $ is
stable, i.e., that it does not depend on time (as already implied in the notation above). One also commonly
assumes that the only time dependence in the laboratory velocity
distribution $f_\lab(\vv,t)$ comes from the motion of the Earth around the
Sun or the daily rotation of the Earth. In other words, one assumes that the WIMP velocity distribution
in the Galactic frame $f_\gal(\uu)$, where $\uu$ is the Galactic WIMP
velocity, is stationary (on the time scale of the experiment).

The laboratory velocity distribution $f_\lab(\vv,t)$ is related to the
Galactic velocity distribution $f_\gal(\uu)$ by a Galilean
transformation:
\begin{align}
f_\lab(\vv,t) = f_\gal\big(\uu\big),
\qquad
\uu=\vv+\vvsun+\vvearth(t) .
\end{align}
Here $\vv$ is the WIMP velocity relative to the laboratory, $\uu$ is the WIMP velocity relative to the  rest frame of the Galaxy, $\vvsun$ is the velocity of the Sun with respect to
the Galactic rest frame, and $\vvearth(t)$ is the velocity of the
Earth with respect to the Sun.\footnote{Since we are interested in the annual modulation, we neglect the daily rotation of the Earth, but our general considerations apply using the velocity of the detector with respect to the Sun in place of $\vvearth(t)$.} A change of integration variables in
Eq.~(\ref{eq:Spt}) gives
\begin{align}
S_{i}(t) = \int \scrH^{\gal}_{i}(\uu,t) \, f_\gal(\uu) \, d^3 u,
\label{eq:Si_of_t}
\end{align}
where
\begin{align}
\scrH^{\gal}_{i}(\uu,t) = \scrH_{i}\big(\uu-\vvsun-\vvearth(t)\big) .
\label{eq:Si_of_t_gal_lab}
\end{align}
Conceptually, this passage to the Galactic frame means that the time dependence of the signal is a property of the motion of the detector in the Galaxy and not of the (Galactic) distribution function. In particular, the characteristics of a modulated signal are ultimately a property of the detector (composition, energy threshold, motion in the Galaxy) and not of the WIMP velocity distribution in the Galactic frame.

The DAMA modulation amplitude is equal to
the coefficient of the $\cos[\omega (t-t_0)]$ term in the
Fourier time-series analysis of the signal. Here $\omega = 2\pi/T$
with the period $T=1$ yr, and the time $t=t_0$ corresponds to the time of maximum modulation. So:
\begin{align}
S_{m,i} = \frac{2}{T} \int_0^{T} dt \, \cos[\omega (t-t_0)] \, S_i(t) .
\end{align}
The unmodulated signal is the time average of the signal over the course of a year,
\begin{align}
S_{0,i} = \frac{1}{T} \int_0^T dt \, S_i(t) .
\end{align}
In the Galactic frame, the time dependence is in the response functions, so one can write
\begin{align}
S_{0,i} & =  \int \scrH^{\gal}_{0,i}(\uu) \, f_\gal(\uu) \, d^3 u ,
\label{eq:S0i}
\\
S_{m,i} & =  \int \scrH^{\gal}_{m,i}(\uu) \, f_\gal(\uu) \, d^3 u ,
\label{eq:Smi}
\end{align}
where
\begin{align}
\scrH^{\gal}_{0,i}(\uu) & = \frac{2}{T} \int_0^{T} dt \, \scrH_i\big(\uu-\vvsun-\vvearth(t)\big),
\label{eq:H0}
\\
\scrH^{\gal}_{m,i}(\uu) & = \frac{2}{T} \int_0^{T} dt \, \cos[\omega (t-t_0)] \, \scrH_i\big(\uu-\vvsun-\vvearth(t)\big).
\label{eq:Hmc}
\end{align}

\section{Constraining the unmodulated signal}

\label{sec:extreme}

Given 
$N$ experimental modulation amplitude measurements $S_{m,j}^{\rm exp}$ ($j=1,\ldots,N$) and their 1$\sigma$
errors $\Delta S_j^{\rm exp}$, and assuming Gaussian fluctuations, the likelihood function ${\cal L}$, function of the set of parameters $\{ S_{m,j} \}$, is given by
\begin{equation} {-2 \ln \cal L}(\{S_{m,j}\})=\sum_{j=1}^{N}\left
    (\frac{S_{m,j}-S_{m,j}^{\rm exp}}{\Delta S_j^{\rm exp}} \right )^2 .
\end{equation}  
Here, for clarity and simplicity, we only show  the dependence of ${\cal L}$ on the expected modulation amplitudes
\begin{align}
S_{m,j} = \int \scrH^{\gal}_{m,j}(\uu) \, f_\gal(\uu) \, d^3u ,
\label{eq:Smkconstraint}
\end{align}
which contain all the dependence on the WIMP velocity distribution function $f_\gal(\uu)$.

\begin{figure}
\begin{center}
\includegraphics[width=0.6\columnwidth]{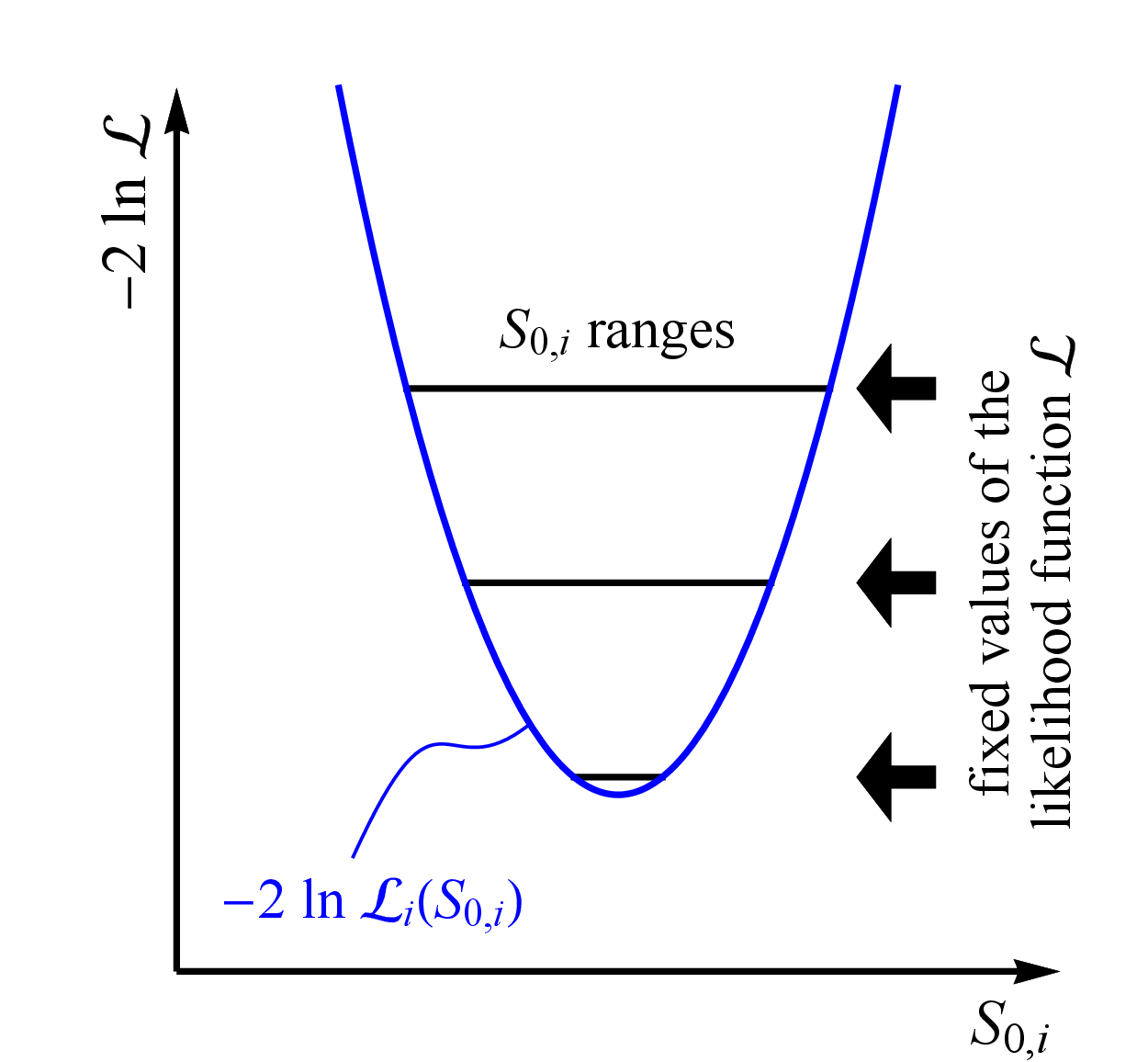}    
\end{center}
\caption{Schematic view of the idea behind the use of the linear optimization
  theorems \cite{representation_theorems,pinelis} to find the
  profile--likelihood of the unmodulated signal rates
  $S_{0,i}$. The profile likelihood ${\cal L}_i(S_{0,i})$ is the maximum value of the likelihood function $\cal L$ at fixed $S_{0,i}$ (parabolic-like line in the figure). The extreme values of $S_{0,i}$ (i.e., its range $[S_{0,i}^{\rm inf}, S_{0,i}^{\rm sup}]$) at a fixed value of the likelihood function $\cal L$ (a horizontal line in the figure) can be found by using extreme distributions (finite sums of streams) for the
  velocity distribution function $f_{\rm gal}(\uu)$. The $S_{0,i}$ ranges at various values of $\cal L$ can be combined to produce the graph of $-2 \ln {\cal L}_i(S_{0,i})$. In practice, we randomly generate extreme distributions with $-2\ln{\cal L} \le -2\ln L_0$, a given number, and fill in the region between a horizontal line and the profile likelihood graph.
  \label{fig:profile_schematic}}
\end{figure}

We are interested in constraining the unmodulated signals
\begin{align}
S_{0,i} = \int \scrH^{\gal}_{0,i}(\uu) \, f_\gal(\uu) \, d^3u, 
\label{eq:S0iconstraint}
\end{align}
in the given energy bins $i=1,\ldots,N$. For this purpose, we construct a joint profile likelihood ${\cal L}_{\rm p}(\{S_{0,i}\})$ for the $S_{0,i}$ ($i=1,\ldots, N$), treating the velocity distribution $f_\gal(\uu)$ as a continuum of nuisance parameters. The joint profile likelihood ${\cal L}_{\rm p}(\{S_{0,i}\})$ is defined as the maximum value of the likelihood function over the set ${\cal A}(\{S_{0,i}\})$ of distribution functions that satisfy Eq.~(\ref{eq:S0iconstraint}),
\begin{align}
{\cal L}_{\rm p}(\{S_{0,i}\}) = \sup_{f_\gal\in {\cal A}(\{S_{0,i}\})} {\cal L}(\{S_{m,j}\}) .
\end{align}
(Technically, we use the notation sup instead of max because in the infinite-dimensional space of distribution functions it is not automatically guaranteed that there is a distribution that achieves the maximum, although this does happen in our case.)

The maximum-likelihood estimator of the $S_{0,i}$'s then follows as the location of the maximum ${\cal L}_{\rm p,max}$ in the joint profile likelihood ${\cal L}_{\rm p}(\{S_{0,i}\})$, and confidence regions for any combination of the $S_{0,i}$'s can be obtained through the usual profile likelihood procedure. For example, the standard error ellipsoid in the $N$-dimensional $S_{0,i}$ parameter space (or `pseudo-ellipsoid' in the case of non-Gaussian likelihoods) can be obtained by the condition 
\begin{align}
-2\Delta\ln {\cal L}_{\rm p}(\{S_{0,i}\}) = 1 ,
\end{align}
where 
\begin{align}
-2\Delta\ln {\cal L}_{\rm p}(\{S_{0,i}\}) =  - 2 \ln {\cal L}_{\rm p}(\{S_{0,i}\}) + 2 \ln {\cal L}_{\rm p,max} .
\end{align}
Similarly, projections of the standard error ellipsoid onto any pair of two variables $S_{0,i}$ and $S_{0,j}$ can be obtained either by simple geometric construction or by using a profile likelihood further profiled over the other $S_{0,i}$ parameters. In Section \ref{sec:analysis} we show an example of such 2-dimensional standard error ellipses.

We are interested in particular in finding confidence intervals on each of the $N$ unmodulated signals $S_{0,i}$. For this purpose, we use the profile likelihood function
\begin{align}
{\cal L}_i(S_{0,i}) = \max_{\substack{S_{0,j}\\j\ne i}} {\cal L}_{\rm p}(\{S_{0,j}\}) ,
\end{align}
which is the joint profile likelihood ${\cal L}_{\rm p}(\{S_{0,i}\})$ further profiled over the $S_{0,j}$ with $j\ne i$, and is thus a function of $S_{0,i}$ only. A 1$\sigma$ confidence interval on a single $S_{0,i}$ can then be obtained through the condition
\begin{align}
-2 \Delta \ln {\cal L}_i(S_{0,i}) \le 1,
\end{align}
where
\begin{align}
-2\Delta\ln {\cal L}_i(S_{0,i}) =  - 2 \ln {\cal L}_i(S_{0,i}) + 2 \ln {\cal L}_{\rm p,max} .
\end{align}
Notice that these 1$\sigma$ confidence intervals, sometime called 1$\sigma$ likelihood intervals, have a 68\% coverage probability in the limit of large samples when the likelihood is well approximated by a Gaussian, but do not necessarily have a coverage probability of 68\% if the likelihood is non-Gaussian.

We examined various ways of computing ${\cal L}_{\rm p}(\{S_{0,i}\})$ and ${\cal L}_i(S_{0,i})$, and we have adopted the following procedure. Since both $S_{0,i}$ and $S_{m,i}$ are functionals of the distribution function $f_{\rm gal}$, we can at least conceptually construct a parametric plot of ${\cal L}(\{S_{m,i}\})$ vs.\ $\{S_{0,i}\}$ by using $f_{\rm gal}$ as the parameter. At each point $\{S_{0,i}\}$ there will be many values of ${\cal L}(\{S_{m,i}\})$, and our goal is to find the maximum of those values, which is ${\cal L}_{\rm p}(\{S_{0,i}\})$. In other words, in this geometrical representation, the joint profile likelihood ${\cal L}_{\rm p}(\{S_{0,i}\})$ is the boundary of all possible values of the likelihood ${\cal L}(\{S_{m,i}\})$ when plotted vs.\ $\{S_{0,i}\}$. In practice we cannot implement an infinite number of functions $f_{\rm gal}$ (we tried discretizing the distribution function but the maximization procedure did not converge). However, we can think of constructing the boundary of the likelihood values by ``rotating the plot by 90 degrees,'' i.e., finding the boundary of the $\{S_{0,i}\}$ that have ${\cal L}(\{S_{m,i}\}) \ge {\cal L}_{\rm p}(\{S_{0,i}\})$. The latter problem can be written as an extremization problem for $\{S_{0,i}\}$ for which powerful mathematical theorems exist that reduce the infinitely-many functions $f_{\rm gal}$ to a finite number of parameters, making the solution attainable in practice.

For clarity, we illustrate our procedure for ${\cal L}_i(S_{0,i})$ only, although we used it for the joint profile likelihood ${\cal L}_{\rm p}(\{S_{0,i}\})$. We write our problem as a linear optimization problem for $\{S_{0,i}\}$ over the distribution functions $f_\gal(\uu)$ subject to the constraint that the  likelihood function ${\cal L}(\{S_{m,j}\})$ is greater than or equal to a given number $L_0$, which we will later vary. Our goal is to find the lower and upper bounds
\begin{align}
& S_{0,i}^{\rm inf}(L_0) = \inf_{f_\gal\in{\cal A}(L_0)} \, \int \scrH^{\gal}_{0,i}(\uu) \, f_\gal(\uu) \, d^3u 
\intertext{and}
& S_{0,i}^{\rm sup}(L_0) = \sup_{f_\gal\in{\cal A}(L_0)} \, \int \scrH^{\gal}_{0,i}(\uu) \, f_\gal(\uu) \, d^3u,
\end{align}
over the set ${\cal A}(L_0)$ of distribution functions that satisfy the constraint
\begin{align}
{\cal L}(\{S_{m,j}\}) \ge L_0 .
\end{align}
Varying $L_0$, we obtain the lines $S_{0,i}^{\rm inf}(L_0)$ and $S_{0,i}^{\rm sup}(L_0)$ in the $S_{0,i}$--${\cal L}$ plane, which we then invert to obtain the profile likelihood ${\cal L}_i(S_{0,i})$ as a function of $S_{0,i}$. Figure~\ref{fig:profile_schematic} is a schematic illustration of our procedure.

To compute $S_{0,i}^{\rm inf}(L_0)$ and $S_{0,i}^{\rm sup}(L_0)$ we notice that the likelihood function is a function of $f_\gal(\uu)$ only through the integrals $S_{m,j}$ in Eq.~(\ref{eq:Smkconstraint}). So we rephrase our goal as the following mathematical problem. 
\begin{align}
& \text{Extremize } S_{0,i} = \int \scrH_{0,i}^{\gal}(\uu) \, f_\gal(\uu) \, d^3u
\label{eq:Mset0}
\intertext{over the set ${\cal A}(L_0)$ of distribution functions $f_\gal(\uu)$ that satisfy the $N+1$ moment conditions}
& \int f_\gal(\uu) \, d^3u = 1 ,
\label{eq:Mset1}
\\
& \int \scrH^{\gal}_{m,j}(\uu) \, f_\gal(\uu) \, d^3u = S_{m,j} \quad (j=1,\ldots,N) ,
\label{eq:Mset2}
\intertext{where the moments $S_{m,j}$ are allowed to vary within the region defined by the likelihood condition}
& {\cal L}(\{S_{m,j}\}) \ge L_0  .
\label{eq:Mset3}
\end{align}
Mathematically, this is an optimization problem of the kind discussed for example in \cite{representation_theorems,pinelis}. Specifically, an optimization problem in which the moment set (i.e., the set of distribution functions that obey the moment conditions) is defined by a subset $C(L_0) \subseteq \mathbb{R}^{N+1}$, namely the subset $C(L_0)$ in which the moments $(1,S_{m,1},\ldots,S_{m,N})$ satisfy ${\cal L}(\{S_{m,j}\})\ge L_0$.

The fundamental theorem in this context \cite{pinelis,representation_theorems} states that $S_{0,i}$ achieves its extreme values $S_{0,i}^{\rm inf}(L_0)$ and $S_{0,i}^{\rm sup}(L_0)$ on the extreme distributions of the moment set, which are sums with positive coefficients of a finite number $K\le N+1$ of Dirac delta functions. Here $N+1$ is the number of moment conditions. More precisely,  the extreme distributions of  the moment set defined by the moment conditions (\ref{eq:Mset1}--\ref{eq:Mset2}) have the form
\begin{align}
f_{\rm e}(\uu) = \sum_{k=1}^{K} \lambda_k \, \delta(\uu - \uu_k) ,
\label{eq:fgalextreme}
\end{align}
where $1\le K \le N+1$,
\begin{align}
&
\sum_{k=1}^{K} \lambda_k \, \scrH^{\gal}_{m,j}(\uu_k) = S_{m,j} \qquad (j=1,\ldots,N),
\label{eq:linear_system1}
\\
&
\sum_{k=1}^{K} \lambda_k = 1 ,
\qquad
\lambda_k > 0 \qquad (k=1,\ldots,K),
\label{eq:linear_system2}
\end{align}
and the $K$ $N+1$-dimensional vectors $(1,\scrH^{\gal}_{m,1}(\uu_k),\ldots,\scrH^{\gal}_{m,N}(\uu_k))$, where the index $k=1,\ldots,K$ specifies the vector, are linearly independent. 

Geometrically, the extreme distributions of the moment set are analogous to the vertices of a polyhedron.  In the finite-dimensional case, the fundamental theorem states that the maximum and minimum values of a linear function defined over a polyhedron are achieved at one or more vertices of the polyhedron, and thus to find these extrema it suffices to compute the value of the linear function on the vertices. In the continuum case, the fundamental theorem states that the extrema of a linear functional of the distribution (in our case, each $S_{0,i}$) are achieved at one or more ``vertices'' of the moment set (i.e., at the extreme distributions), and thus to find these extrema it suffices to compute the value of the linear functional on the extreme distributions. The computational advantage is that the fundamental theorem reduces an extremization problem in infinite dimensions (the moment set) into an extremization problem in a finite number of dimensions (the space of extreme distributions, which has dimension at most $(1+d) N'$, where $N'=N+1$ is the number of moment conditions and $d$ is the dimensionality of the velocity space). 

Physically, each delta-function distribution $ \delta(\uu - \uu_k) $ in the expression of an extreme distribution, Eq.~(\ref{eq:fgalextreme}), represents a stream of velocity $\uu_k$ and zero velocity dispersion. An extreme distribution is a weighted average of streams with weights $\lambda_k$. The fundamental theorem allows the computation of $S_{0,i}^{\rm inf}(L_0)$ and $S_{0,i}^{\rm sup}(L_0)$ by parametrizing the velocity distribution as a weighted average of no more than $N+1$ streams in velocity space, where $N+1$ is the number of moment conditions, including the normalization condition.\footnote{The fundamental theorem is also the rigorous mathematics behind the ``interesting'' facts that the number of steps in $\tilde{\eta}(\vmin)$ is less than or equal to the number of bins for a binned likelihood \cite{1403.4606}, or that $N_O$ streams suffice to minimize an unbinned likelihood with $N_O$ events \cite{1403.6830,1502.07682}, or that only $1+p+q$ grid points have a nonzero distribution function in the presence of $1+p+q$ moment conditions \cite{IbarraRappeltMIAPP}.}

The fundamental theorem translates the mathematical problem (\ref{eq:Mset0}--\ref{eq:Mset3}) into the following one.
\begin{align}
& \text{Extremize } && S_{0,i} = \sum_{k=1}^{K} \lambda_k \, \scrH^{\gal}_{0,i}(\uu_k) 
\label{eq:sysK1}
\\
& \text{over $\lambda_k$ and $\uu_k$ subject to } &&
1\le K \le N+1,
\label{eq:sysK2}
\\[1ex]
& &&
\lambda_k > 0 \qquad (k=1,\ldots,K), 
\label{eq:sysK3}
\\
& &&
\sum_{k=1}^{K} \lambda_k = 1 ,
\label{eq:sysK4}
\\
& &&
S_{m,j}  = \sum_{k=1}^{K} \lambda_k \, \scrH^{\gal}_{m,j}(\uu_k) \qquad (j=1,\ldots,N),
\label{eq:sysK5}
\\[1ex]
& &&
{\cal L}(\{S_{m,j}\}) \ge L_0 .
\label{eq:sysK6}
\end{align}

In practice this means that at fixed $K$ and given $L_0$, the maximal range of the
$S_{0,i}$ integral computed using Eq.~(\ref{eq:sysK1}) is swept by the $\lambda_k$, $\uu_k$ parameters that
satisfy the constraints (\ref{eq:sysK3}-\ref{eq:sysK6}). The full range of $S_{0,i}$ is then obtained by combining the $N+1$
intervals for $1\le K \le N+1$.

It is very important to understand that this method does not in general give the optimal velocity distribution, or the maximum-likelihood velocity distribution.
In fact, given a value $L_0$ of the likelihood, there are in general many $S_{m,i}$ that have the same likelihood (all those on the likelihood contour level ${\cal L}(\{S_{m,i}\})=L_0$). But even if there is only one set of $S_{m,i}$ that corresponds to a given value of the likelihood (and this happens at the point of absolute maximum likelihood for concave likelihood functions), there are in general many velocity distributions with the same moments $S_{m,i}$: some of them are extreme distributions (sums of streams), and some are continuous distributions, or more precisely, continuous linear combinations of sums of streams of the form
\begin{align}
f(\uu) = \int_0^1 \sum_{k=1}^{K} \lambda_k(\alpha) \, \delta\big( \uu - \uu_k(\alpha) \big) \, d\alpha.
\label{eq:degeneratef}
\end{align}
In particular, although the value of the maximum likelihood can be obtained using only sums of streams, there is in general an infinite number of distributions, some discrete and some continuous, that maximize the likelihood. So even if we use extreme distributions to find the extreme values of $S_{0,i}$, it is not correct to think that in general these sums of streams are the only velocity distributions giving those extreme values. The reason we can use the methods of this Section to estimate the unmodulated signal is that we are not interested in finding the optimal velocity distribution but in performing a maximum-profile-likelihood analysis of quantities like $S_{0,i}$ that are integrals of the velocity distribution. For this task, the method described in this Section is adequate and mathematically sound.

\section{Analysis: isotropic case}
\label{sec:analysis}

In this Section, we apply the general method described in Section~\ref{sec:extreme} to a specific case: a halo-independent estimate of the unmodulated DAMA signal. Since this is the first implementation of our method, we have made some simplifying assumptions that have allowed us to explore and understand the difficulties and merits of the method itself. First and foremost, to reduce the computing time, we have restricted our analysis to WIMP velocity distributions that are isotropic in the Galactic reference frame, i.e., $f_{\rm gal}(\uu)=f_{\rm gal}(u)$ with $u=|\uu|$, so that the distribution functions depend on one variable only (the magnitude $u$) instead of three (the components of $\uu$). Under this assumption, the Galactic response functions can be replaced by the angle-averaged Galactic response functions, defined by an average over the directions of the vector $\uu$ as
\begin{align}
\overline{\scrH}_{0,i}^{\gal}(u) & = \frac{1}{4\pi} \int \scrH^{\gal}_{0,i}(\uu)  \, d\Omega_u ,
\\
\overline{\scrH}_{m,i}^{\gal}(u) & = \frac{1}{4\pi} \int \scrH^{\gal}_{m,i}(\uu)  \, d\Omega_u .
\end{align}
Then our extremization problem reads
\begin{align}
& \text{Extremize } && S_{0,i} = \sum_{k=1}^{K} \lambda_k \, \overline{\scrH}^{\rm gal}_{0,i}(u_k) 
\label{eq:sysKI1}
\\
& \text{over $\lambda_k$ and $u_k$ subject to } &&
1\le K \le N+1,
\label{eq:sysKI2}
\\[1ex]
& &&
\lambda_k > 0 \qquad (k=1,\ldots,K), 
\label{eq:sysKI3}
\\
& &&
\sum_{k=1}^{K} \lambda_k = 1 ,
\label{eq:sysKI4}
\\
& &&
S_{m,j}  = \sum_{k=1}^{K} \lambda_k \, \overline{\scrH}^{\rm gal}_{0,i}(u_k) \qquad (j=1,\ldots,N),
\label{eq:sysKI5}
\\[1ex]
& &&
{\cal L}(\{S_{m,j}\}) \ge L_0 .
\label{eq:sysKI6}
\end{align}
Moreover, the isotropic extreme distribution functions are
\begin{align}
\overline{f}_{\rm e}(u) = \sum_{k=1}^{K} \lambda_k \, \delta(u - u_k) ,
\end{align}
where $\overline{f}_{\rm e}(u) = 4 \pi u^2 f_{\rm e}(u) $.
While mathematically appropriate and well-defined, physically these isotropic extreme distributions do not describe a collection of streams in velocity space but rather some sort of spherical shells in velocity space.

An additional simplifying assumption we make in this first application of our method is to consider only spin-independent scattering off sodium in the DAMA NaI detector. Thus we restrict our analysis to light
WIMP masses ($m_{\chi}\le$ 15 GeV) for which WIMP elastic scattering
off iodine is below threshold (for a constant iodine quenching factor $Q_I=0.09$ and Galactic escape speeds less than $\sim\!\!580$~km/s). It must be noted that the results of our analysis apply also to cross sections that are not spin-independent but have a mild energy dependence in the 2--4 keVee energy range.

For our analysis we use the $N=12$ DAMA cosine modulation amplitudes
in the lowest energy bins in Fig.~8 of Ref.~\cite{dama}. We list them
in Table~\ref{tab:S}. These measurements were obtained using a total
exposure of 1.33 ton yr. The signal is concentrated in the first 6
bins, and the other 6 bins act as a control set with no modulation
signal. Data are also available for the sine modulation
amplitudes~\cite{dama} but under our simplifying assumption of
isotropic velocity distribution, the sine modulation response
functions vanish identically (see Appendix), and thus including them
in the likelihood would amount to adding an irrelevant constant. The
DAMA Collaboration published also time series of its modulation data
\cite{dama}, with time binnings ranging from ~ 30 days (close to the
maxima and minima of the oscillation) and ~70 days (close to its
equilibrum points). However it has been shown that the corresponding
error bars can easily accommodate sizeable distortions from a
sinusoidal time dependence of the signal (see for instance the
discussion in Section 5 of Ref.\cite{1512.00593}, relative to the case
of a Maxwellian distribution yielding modulation fractions of order
unity when the incoming WIMP velocities are very close to the escape
velocity), so the ensuing constraint has no impact on our analysis and
we neglect it.

For the DAMA response functions off sodium, we take $i$ to be the index of the energy bin in the
electron--equivalent energy $E_{ee}$. The latter is related to the recoil energy
$E_R$ on average by $\overline{E}_{ee}=Q(E_R) \, E_R$, where $Q(E_R)$ is the quenching factor, with an additional smearing due to a finite energy resolution. We use the DAMA response functions for elastic spin-independent WIMP--nucleus scattering, but our results extend practically unchanged to other elastic interactions in which the velocity and/or energy dependence of the cross section is negligible in the DAMA energy bins (e.g., spin-dependent interactions or other nonrelativistic effective operators that do not show a large variation with recoil energy or WIMP velocity). The DAMA response function for the $i$-th bin with electron--equivalent energies in the range $E_{ee,i} \le E_{ee} \le E_{ee,i+1}$ is 
 \cite{del_nobile_generalized}:
\begin{align}
  & \scrH_{i}(v) = \frac{N_T}{M_{\rm det} \, \Delta E} \, \frac{\rho_{\chi}}{m_{\chi}} \, \sigma_{\chi T} \, \widehat{\scrH}_{i}(v),\label{eq:earth_response_function}\\ 
  &  \widehat{\scrH}_{i}(v) = \frac{v}{E_R^{\rm max}(v)} \int_0^{E_R^{\rm max}(v)}\, dE_R \, F\big(E_R,v\big) 
  \int_{E_{ee,i}}^{E_{ee,i+1}} dE_{ee}\,
  \scrG_T(E_{ee},E_R)\, \epsilon(E_{ee}).
\label{eq:earth_response_function_hat}
\end{align}
Here $N_T$ is the number of target nuclei in the detector, $M_{\rm det}$ is the mass of the detector, $\Delta E = E_{ee,i+1}-E_{ee,i}$ is the width of the energy bins, 
$\rho_\chi$ is the WIMP density in the neighborhood of the Sun, $\sigma_{\chi T}$ is
a reference cross section representing the strength of the WIMP--nucleus interaction, 
\begin{align}
F(E_R,v) = \frac{E_R^{\rm max}(v)}{\sigma_{\chi T}} \, \frac{d\sigma_{\chi T}}{dE_R}
\end{align}
is a form factor which depends on the assumed interaction operator, $\scrG_T(E_{ee},E_R)$ is the energy resolution
smearing function, $\epsilon(E_{ee})$ is an acceptance function
including the effect of experimental cuts, and
\begin{align}
E_R^{\rm max}(v)=\frac{2\mu_{\chi T}^2 v^2}{m_T}
\label{eq:ERmax}
\end{align}
is the maximum recoil energy achievable for a WIMP of speed $v$ scattering off a nucleus of mass $m_T$ (here $\mu_{\chi T}=m_\chi m_T/(m_\chi+m_T)$ is the reduced WIMP--nucleus mass).

The reduced response function  $\widehat{\scrH}_{i}(v)$ has dimensions of  velocity. The dimensionless ratio  $\widehat{\scrH}_{i}(v)/v$ has the immediate  physical interpretation as the fraction of an incoming monochromatic  flux of speed $v$ that is detected in the $i$-th electron--equivalent energy bin.  

 We take $\scrG_T(E_{ee},E_R)$ to be a Gaussian in $E_{ee}$ centered at $\overline{E}_{ee} = Q(E_R) \, E_R$ and with width $\sigma_{\rm rms}/\mbox{keVee}=0.0091
\, (\overline{E}_{ee}/\mbox{keVee})+0.448 \, (\overline{E}_{ee}/\mbox{keVee})^{1/2}$. We further assume that $\scrG_T(E_{ee},E_R)$ vanishes below the
hardware threshold of 1 keVee. We assume
a constant quenching factor $Q(E_R)=0.3$ for sodium. 

For the form factor $F(E_R,v)$ we use the
spin-independent form factor of ${}^{23}$Na as given by the Helm form
in \cite{lewinsmith}, in correspondence of which $\sigma_{\chi T}$ is the point--like ${}^{23}$Na--WIMP cross section. For the WIMP masses we consider
($m_\chi\lesssim 15~{\rm GeV}$), this form factor varies negligibly in our
analysis: by less than
1\% over the 2--4 keVee range where the DAMA modulation is significant,
and by $\lesssim 3$\% over the whole 2--8 keVee range. By the same
token, our analysis applies to all cases in which the variation of the
${}^{23}$Na form factor $F(E_R,v)$ is negligible. Notice in
addition that for such ${}^{23}$Na form factors, any difference in
strength between WIMP--proton and WIMP--neutron interactions can be
included in the reference cross section $\sigma_{\chi T}$. Thus our
analysis applies equally well to ${}^{23}$Na--WIMP elastic scattering
that is, for example, any combination of isoscalar and isovector
spin-independent interactions, any combination of spin-dependent
interactions (for which the ${}^{23}$Na form factors vary by $\lesssim 1$\%
over the whole 2--8 keVee range), and so on.

\begin{figure}[t]
\begin{center}
\includegraphics[width=0.8\columnwidth]{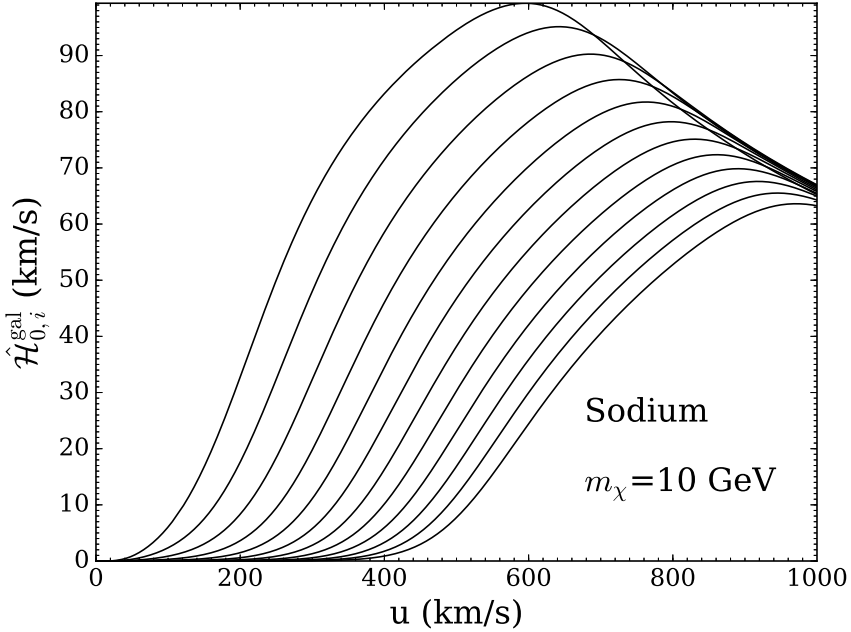}
\end{center}
\caption{Unmodulated angle-averaged Galactic DAMA response functions $\widehat{\scrH}^{\gal}_{0,i}(u)$, calculated using Eqs.~(\ref{eq:earth_response_function_hat}) and~(\ref{eq:h0_fourier}), for elastic spin-independent WIMP scattering off
  sodium targets and $m_{\chi}=10~{\rm GeV}$. From left to right, each curve
  corresponds to one of the first 12 energy bins of 0.5 keVee width counting from the low-energy 
  threshold ($2~{\rm keVee} \le E_{ee}\le 8~{\rm keVee}$).}
\label{fig:response_functions_h0_mchi_10}
\end{figure}

\begin{figure}[t]
\begin{center}
\includegraphics[width=0.8\columnwidth]{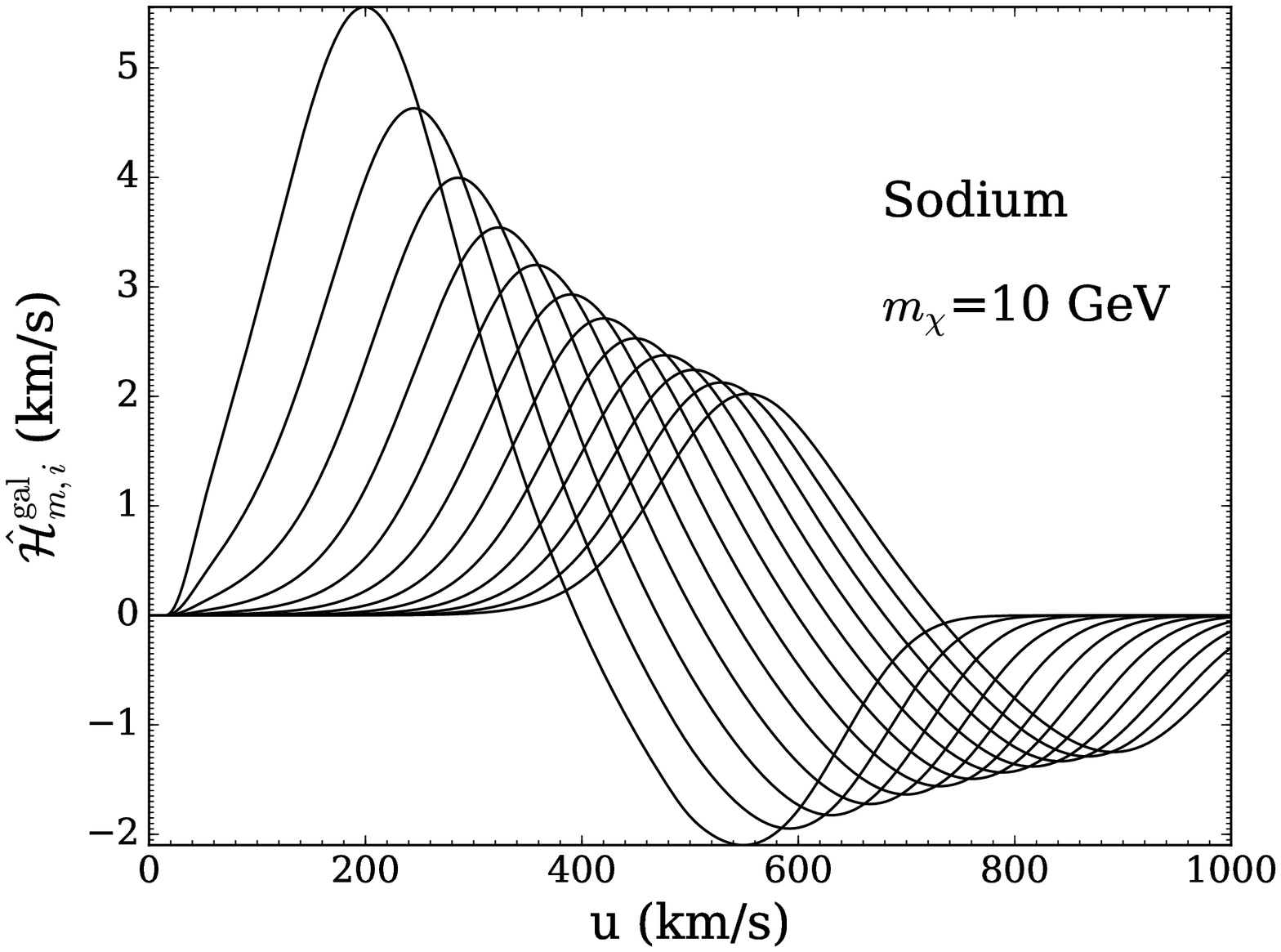}
\end{center}
\caption{Same as Fig.~\ref{fig:response_functions_h0_mchi_10} but
  for the modulated angle-averaged Galactic response functions $\widehat{\scrH}^{\gal}_{m,i}(u)$, calculated
  using Eqs.~(\ref{eq:earth_response_function_hat}) and~(\ref{eq:hm_fourier}).}
\label{fig:response_functions_hm_mchi_10}
\end{figure}

We implement the angle-averaged Galactic
response functions $\overline{\scrH}^{\rm gal}_{0,i}(u)$
and $\overline{\scrH}^{\rm gal}_{m,i}(u)$ for elastic spin-independent 
WIMP--sodium scattering according to
Eqs.~(\ref{eq:h0_fourier}--\ref{eq:hm_fourier}) in the Appendix, with
$\scrH_{i}(v)$ given by
Eq.~(\ref{eq:earth_response_function}) and $\vsun=232$~km/s. Figs.~\ref{fig:response_functions_h0_mchi_10} and~\ref{fig:response_functions_hm_mchi_10} show these response functions for $m_\chi = 10 ~{\rm GeV}$. In these figures, we plot the reduced response functions 
\begin{align}
\widehat{\scrH}^{\gal}_{0,i}(u) & = \frac{m_\chi M_{\rm det} \, \Delta E }{N_T \rho_{\chi}\sigma_{\chi T}} \, \, \overline{\scrH}^{\gal}_{0,i}(u),
\\
\widehat{\scrH}^{\gal}_{m,i}(u) & = \frac{m_\chi M_{\rm det} \, \Delta E }{N_T \rho_{\chi}\sigma_{\chi T}} \, \, \overline{\scrH}^{\gal}_{m,i}(u).
\end{align}
The reduced response functions have dimensions of velocity and represent the Galactic
response functions normalized in the same way as $\widehat{\scrH}_i(v)$ in
Eq.~(\ref{eq:earth_response_function_hat}). Conceptually, the reduced response functions divided by $u$ give the fraction of WIMPs of Galactic speed $u$ that contribute to the detectable event rate.

As anticipated, we cast the
modulation effect as a property of the detector Galactic response
function: the
only information needed to obtain the reduced Galactic response functions for a given WIMP-nucleus cross section, both modulated and unmodulated, are the experimental
properties of the detector and the motion of the detector in the Galaxy.

Once the Galactic response functions are given, the procedure outlined
in Section \ref{sec:extreme}, namely solving the extremization
problem~(\ref{eq:sysK1}--\ref{eq:sysK6}), can be used to estimate the unmodulated signals $S_{0,i}$ from the DAMA data on $S_{m,i}$. The following three
considerations are relevant in actually implementing the method
of Section \ref{sec:extreme}.

(1) Although no assumptions on $f_{\rm gal}(u)$ are needed in the extremization problem~(\ref{eq:sysK1}--\ref{eq:sysK6}), it appears  natural to assume that there is a maximum speed for the WIMPs in the Galaxy. Thus we assume that $f_{\rm gal}(u)$ vanishes when $u$ exceeds a maximum velocity $u_{\esc}$, which we take to be the escape speed from the Galaxy as quoted in \cite{vesc,vesc_salucci}, 
\begin{align}
u_{\esc}=550~{\rm km/s}.
\end{align}
(This value is within the updated measurements in~\cite{vescnew}.) We therefore restrict $u\le u_{\esc}$.

(2) As seen in Eq. (\ref{eq:earth_response_function}), and as true in general, the absolute normalization of the response functions contains the unknown factor $\rho_{\chi} \, \sigma_{\chi T}$. However, its value cancels out in the determination of the $S_{0,i}$ from the $S_{m,i}$, essentially because the same factor $\rho_\chi \, \sigma_{\chi T}$ appears in both. In principle, one could fix the value of $\rho_\chi \sigma_{\chi T}$ that appears in the nonreduced response functions $\overline{\scrH}^{\gal}_{0,i}(u)$ and $\overline{\scrH}^{\gal}_{m,i}(u)$ in the extremization problem~(\ref{eq:sysK1}--\ref{eq:sysK6}), solve it as it stands, and then combine the solutions as the value of $\rho_\chi \sigma_{\chi T}$ is varied from zero to infinity. Alternatively, and this is the procedure we actually implement, one can use reduced response functions $\widehat{\scrH}^{\gal}_{0,i}(u)$ and $\widehat{\scrH}^{\gal}_{m,i}(u)$, which do not contain the factor $\rho_\chi \sigma_{\chi T}$, rescale the coefficients $\lambda_k$ to 
\begin{align}
\widehat{\lambda}_k = \frac{N_T \rho_{\chi}\sigma_{\chi T}}{m_\chi M_{\rm det} \, \Delta E } \, \,  \lambda_k ,
\end{align}
so that 
\begin{align}
\lambda_k \, \overline{\scrH}^{\gal}_{0,i}(u_k) = \widehat{\lambda}_k \widehat{\scrH}^{\gal}_{0,i}(u_k) ,
\qquad 
\lambda_k \, \overline{\scrH}^{\gal}_{m,i}(\uu_k) = \widehat{\lambda}_k \widehat{\scrH}^{\gal}_{m,i}(u_k) ,
\end{align}
and solve the modified extremization problem in which the normalization condition $\sum_{k=1}^{K} \lambda_k = 1$ is replaced by the condition
\begin{align}
\sum_{k=1}^{K} \widehat{\lambda}_k > 0 .
\end{align}
Since this condition is already contained in the conditions $\widehat{\lambda}_k > 0$, one of the moment conditions effectively disappears, and the extreme values of the $S_{0,i}$ can be found with sums of up to $N$, instead of $N+1$, streams in velocity space, provided the extreme distributions contain the rescaled coefficients $\widehat{\lambda}_k$ instead of $\lambda_k$.

(3) Having dropped the normalization condition on the $\lambda_k$ as described in the previous paragraph, the $K$ streams ($1\le K \le N$) of an extreme distribution must have speeds $u_k$ such that the $K$ $N$-dimensional vectors $(\widehat{\scrH}_{m,1}(u_k),\ldots,\widehat{\scrH}_{m,N}(u_k))$ are linearly independent. Now the experimental threshold in observed energy (1 keVee in our treatment of DAMA; see our discussion of ${\cal G}_T(E_{ee},E_R)$ after Eq.~(\ref{eq:earth_response_function_hat})) induces a region below threshold in velocity space, comprised of all speeds $u$ for which the response functions vanish simultaneously, $\widehat{\scrH}_{m,i}(u)=0$ ($i=1,\ldots,N$) for $u$ below threshold. For example, in the isotropic case we consider, the threshold speed for the modulated Galactic response functions turns out to be
\begin{align}
u_{\thr} = 210.13,\, 30.91,\, 0~{\rm km/s} \text{ for } m_\chi=5, 10, 15~{\rm GeV, respectively.}
\end{align}
Thus, if the velocity of one or more of the $K$ streams is below threshold, the $K$ vectors mentioned above are not linearly independent (one or more of them is the zero vector). On the other hand, the streams with velocity below threshold do not contribute to the $S_{m,i}$ signals at all (indeed, all the response functions are zero for these streams).  Thus a sum of $K$ streams in which some streams are below threshold is effectively an extreme distribution with less than $K$ streams. Since we let $K$ vary from 1 to $N$, it is obvious that it is enough for extreme distributions to include only streams above threshold. Therefore we allow only $u > u_{\thr}$.

The practical implementation of the method described above is
conceptually quite simple, although the use of the parametrization
(\ref{eq:fgalextreme}) for the extreme distributions of the moment set
requires to explore a parameter space of large dimensionality ($2N=24$
in our $N=12$ case with isotropic Galactic velocity distribution;
$4N=48$ if we explored anisotropic Galactic velocity distributions).
This kind of task is efficiently performed by using the technique of
Markov chains, which makes use of the likelihood function itself to
optimize the sampling procedure.\footnote{Alternatively one could use
  grids in velocity space and increase the grid resolution to increase
  the precision of the computed extreme values, or one could use algorithms to directly maximize and minimize $S_{0,i}$ under the given constraints. We tried some of those methods without success. We chose to generate Monte-Carlo samples of the likelihood to gain confidence on our method.} To this aim we use
the Markov--Chain Monte Carlo (MCMC) code emcee~\cite{emcee} to generate a
large number of sets $\{ u , \widehat{\lambda} \} = \{ u_1, \ldots,
u_K, \widehat{\lambda}_1, \ldots, \widehat{\lambda}_K \} $ of Galactic
speeds $u_k$ and coefficients $\widehat{\lambda}_k$ with $1\le k \le
K$, $1\le K \le N$, $u_{\thr} < u_k \le u_{\esc}$ and
$\widehat{\lambda}_k > 0$. For each value of $K=1,...,N$, we generate
a Markov chain of $5\times10^6$ points using 250 independent walkers
and a standard Metropolis-Hastings sampler.
  
\begin{figure}[!h]
\begin{center}
\includegraphics[width=0.99\columnwidth, bb=40 81 796 1192, height=18cm]{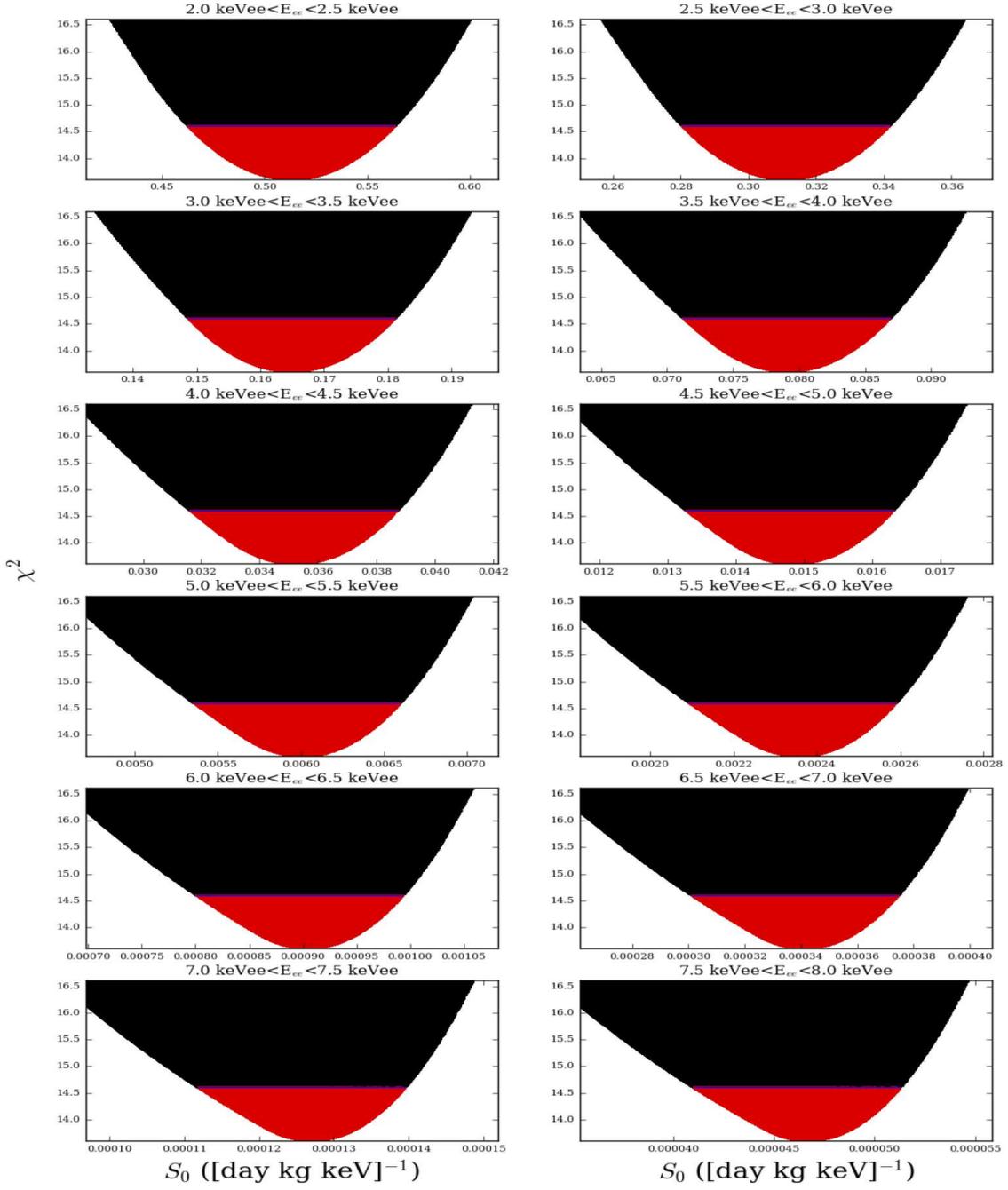}    
\end{center}
\caption{Scatter plots of $\chi^2=-2 \ln ({\cal L})$ vs.\
  $S_{0,i}$ ($i=1,\ldots,12$) for the WIMP mass $m_\chi=5 ~{\rm GeV}$. Each panel corresponds to one of
  the DAMA energy bins between 2 keVee and 8 keVee.  The boundary of the region covered by points is the graph of $-2\ln {\cal L}_i(S_{0,i})$, where ${\cal L}_i(S_{0,i})$ is the profile likelihood for $S_{0,i}$. The points colored in red have $\chi_i^2\le\chi^2_{i,\rm min}+1$, with
  $\chi^2_{i,\rm min}$ the absolute minimum of the $\chi_i^2$, and determine
  the 1$\sigma$ confidence intervals of $S_{0,i}$.}
\label{fig:mchi_5_profile}
\end{figure}

\begin{figure}[!h]
\begin{center}
\includegraphics[width=0.99\columnwidth, bb=40 81 796 1192, height=18cm]{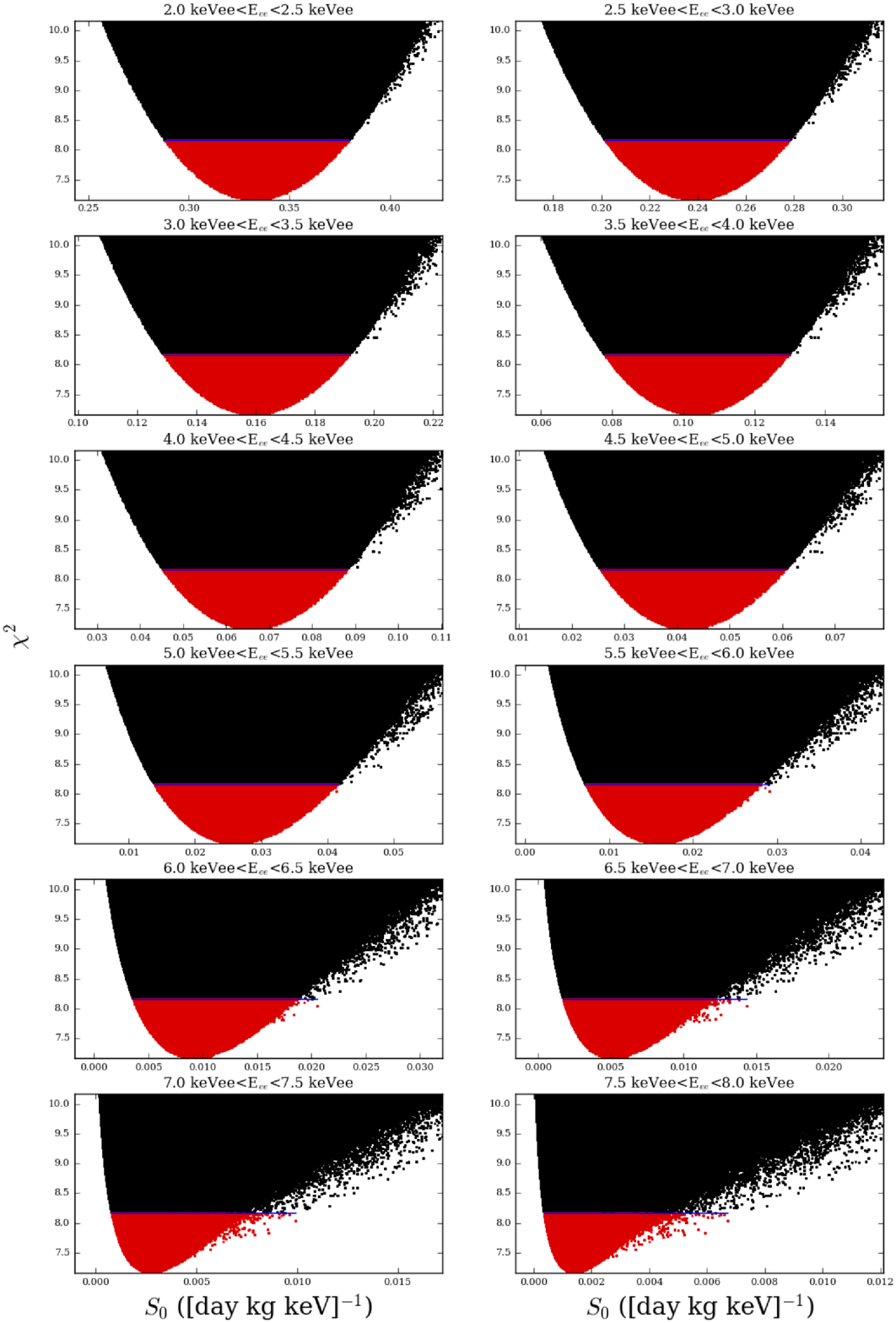}    
\end{center}
\caption{Same as Fig.~\ref{fig:mchi_5_profile} but for $m_{\chi}=10~{\rm GeV}$.}
\label{fig:mchi_10_profile}
\end{figure}

\begin{figure}[!h]
\begin{center}
\includegraphics[width=0.99\columnwidth, ,bb=40 81 796 1192, height=18cm]{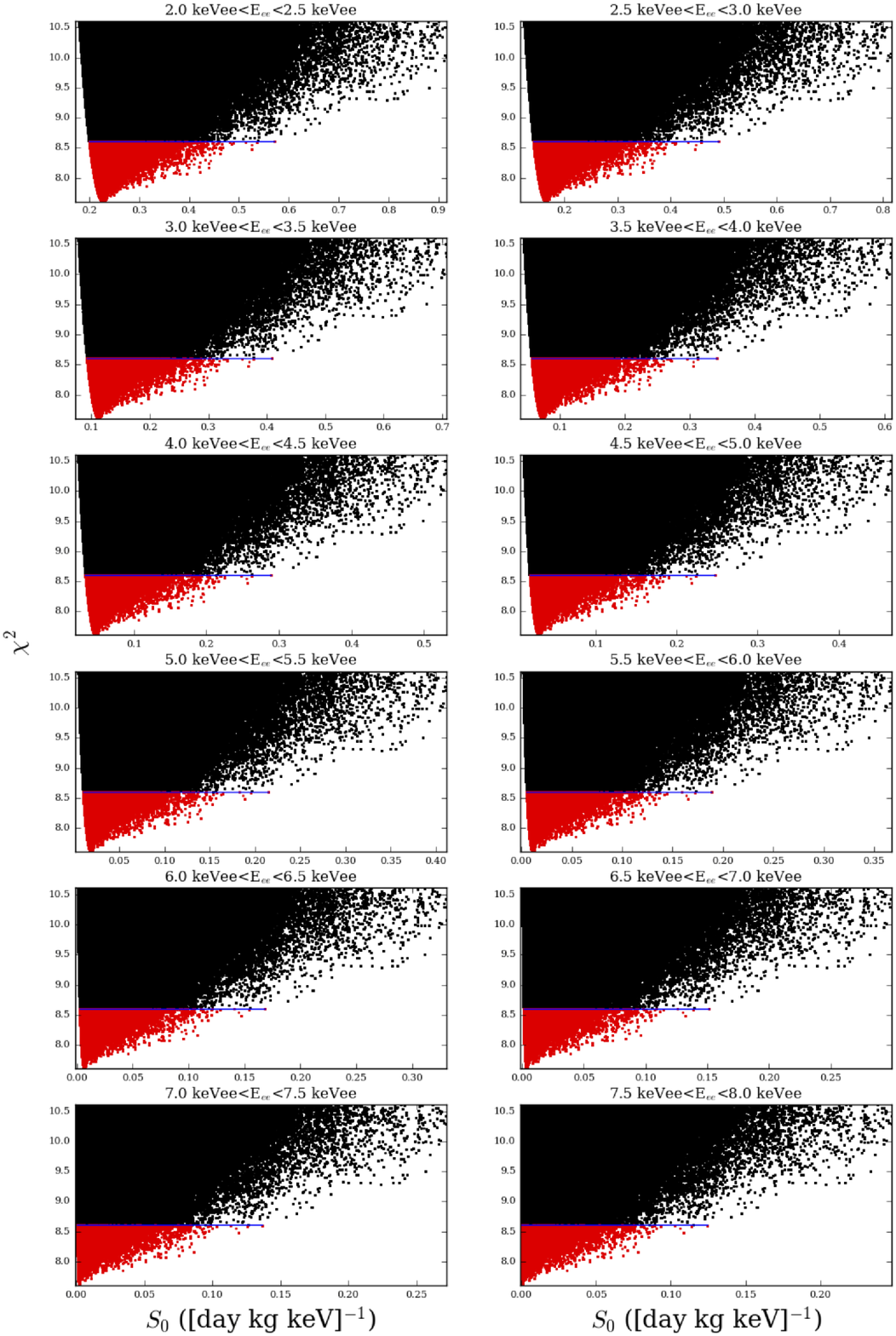}    
\end{center}
\caption{Same as Fig.~\ref{fig:mchi_5_profile} but for $m_{\chi}=15~{\rm GeV}$.}
\label{fig:mchi_15_profile}
\end{figure}

For each MCMC-generated set $\{ u, \widehat{\lambda} \} $ we calculate
both $\chi^2 = -2 \ln {\cal L}$ and $S_{0,i}$ for $i=1,\ldots, N$
$(N=12)$ and produce $N$ scatter plots of $\chi^2$ vs.\ each of the
$S_{0,i}$.  The resulting $(\chi^2,S_{0,i})$ scatter plots are shown
in Figs.~\ref{fig:mchi_5_profile}, \ref{fig:mchi_10_profile}
and~\ref{fig:mchi_15_profile} for three representative values of the
WIMP mass: $m_{\chi}=5~{\rm GeV}$, 10 GeV, and 15 GeV,
respectively. In each of these figures, the 12 panels correspond to
one of the 12 DAMA energy bins between 2 keVee and 8 keVee.  In each
panel, the boundary of the region covered by points gives the profile
likelihood for the $S_{0,i}$ in that panel. More precisely, the
boundary in the $i$-th panel is the graph of $\chi_i^2 = -2\ln {\cal
  L}_i(S_{0,i})$, where ${\cal L}_i(S_{0,i})$ is the profile
likelihood for $S_{0,i}$ with the velocity distribution treated as a
continuum of nuisance parameters. We use two colors for the points,
red and black. The points in red (black) have
$\chi^2_i\le\chi^2_{i,\rm min}+1$ ($\chi^2_i>\chi^2_{i,\rm min}+1$),
where $\chi^2_{i,\rm min}$ is the minimum of the $\chi^2_i$. The value
of $S_{0,i}$ where $\chi^2_i$ is minimum is the maximum-likelihood
estimate of $S_{0,i}$. The lowest and highest values of $S_{0,i}$ in
the $\chi^2_i\le\chi^2_{i,\rm min}+1$ region (points in red, which
occur at the top boundary of the red region) are the endpoints of the
1$\sigma$ confidence interval for $S_{0,i}$. 

\begin{figure}[t]
\begin{center}
\includegraphics[width=0.8\columnwidth]{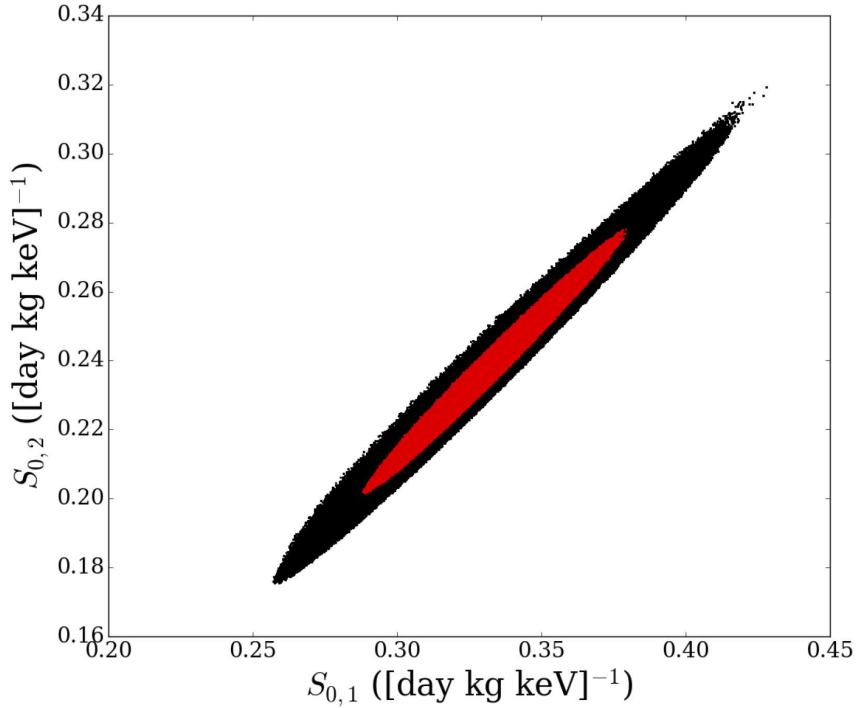}
\end{center}
\caption{The regions $-2\Delta \ln {\cal L}_{\rm p}(\{S_{0,i}\})\le1$ (red points; standard error ellipse) and $-2\Delta \ln {\cal L}_{\rm p}(\{S_{0,i}\})\le3$ (black points) in the plane $S_{0,1}$--$S_{0,2}$ of the unmodulated signals in the first and second energy bins for $m_\chi=10$ GeV. }
\label{fig:ellipse}
\end{figure}

To illustrate that our analysis includes the correlations among  the $S_{0,i}$'s, in Fig.~\ref{fig:ellipse} we show the regions $-2\Delta \ln {\cal L}_{\rm p}(\{S_{0,i}\})\le1$ (red points) and $-2\Delta \ln {\cal L}_{\rm p}(\{S_{0,i}\})\le3$ (black points) in the plane $S_{0,1}$--$S_{0,2}$ of the unmodulated signals in the first and second energy bins for $m_\chi=10$ GeV. The 1$\sigma$ confidence intervals on $S_{0,1}$, for example, are obtained by projecting the red ellipse onto the $S_{0,1}$ axis.

\begin{figure}[t]
\begin{center}
\includegraphics[width=0.8\columnwidth]{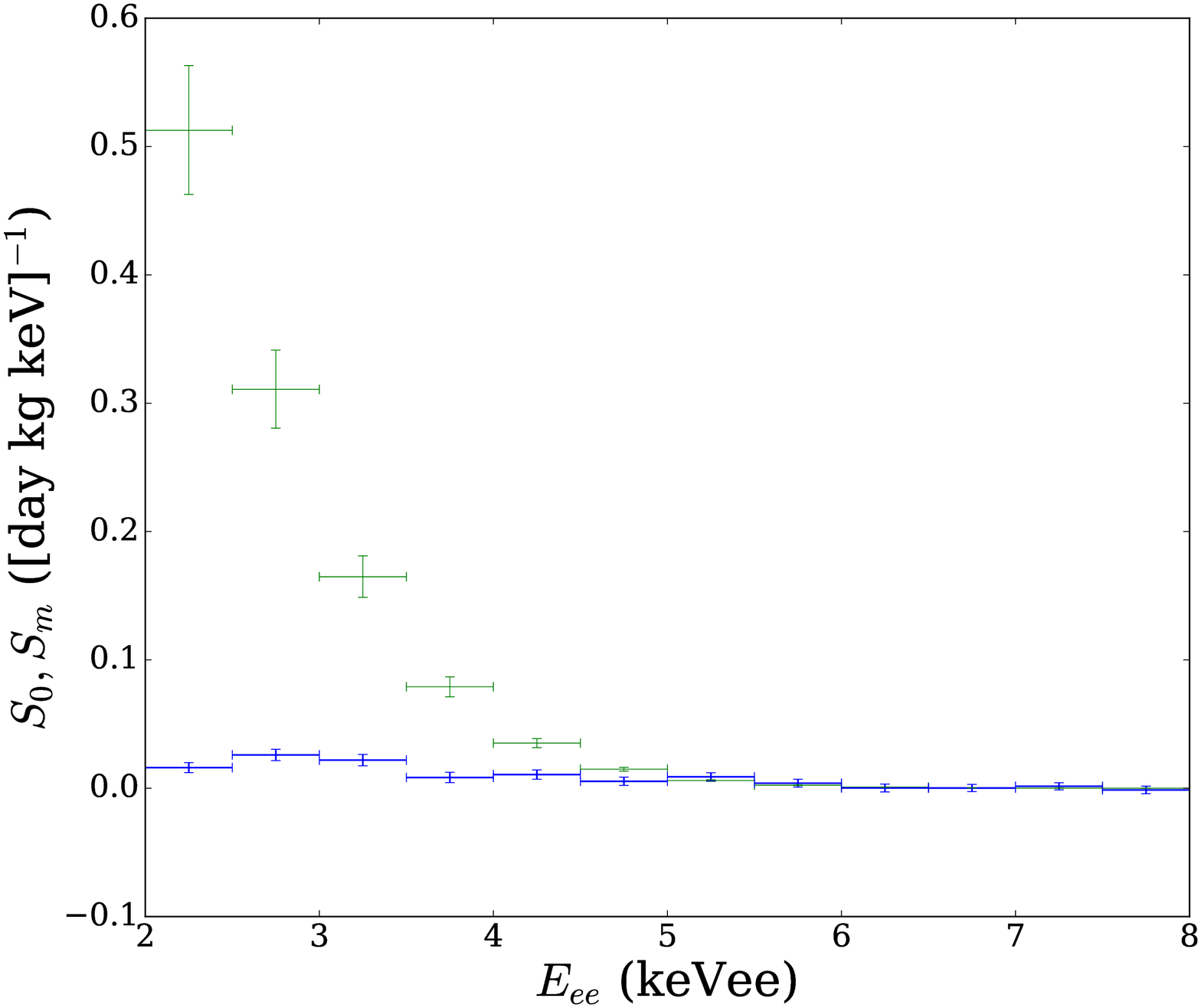}
\end{center}
\caption{Maximum-likelihood estimates and 1$\sigma$ confidence intervals of $S_{0,i}$ ($i=1,\ldots,12$) 
for $m_{\chi}=5$ GeV, obtained
  from the scatter plots in Fig.~\ref{fig:mchi_5_profile}, versus the
  electron--equivalent energy $E_{ee}$.}
\label{fig:mchi_5_e_s0}
\end{figure}

\begin{figure}[t]
\begin{center}
\includegraphics[width=0.8\columnwidth]{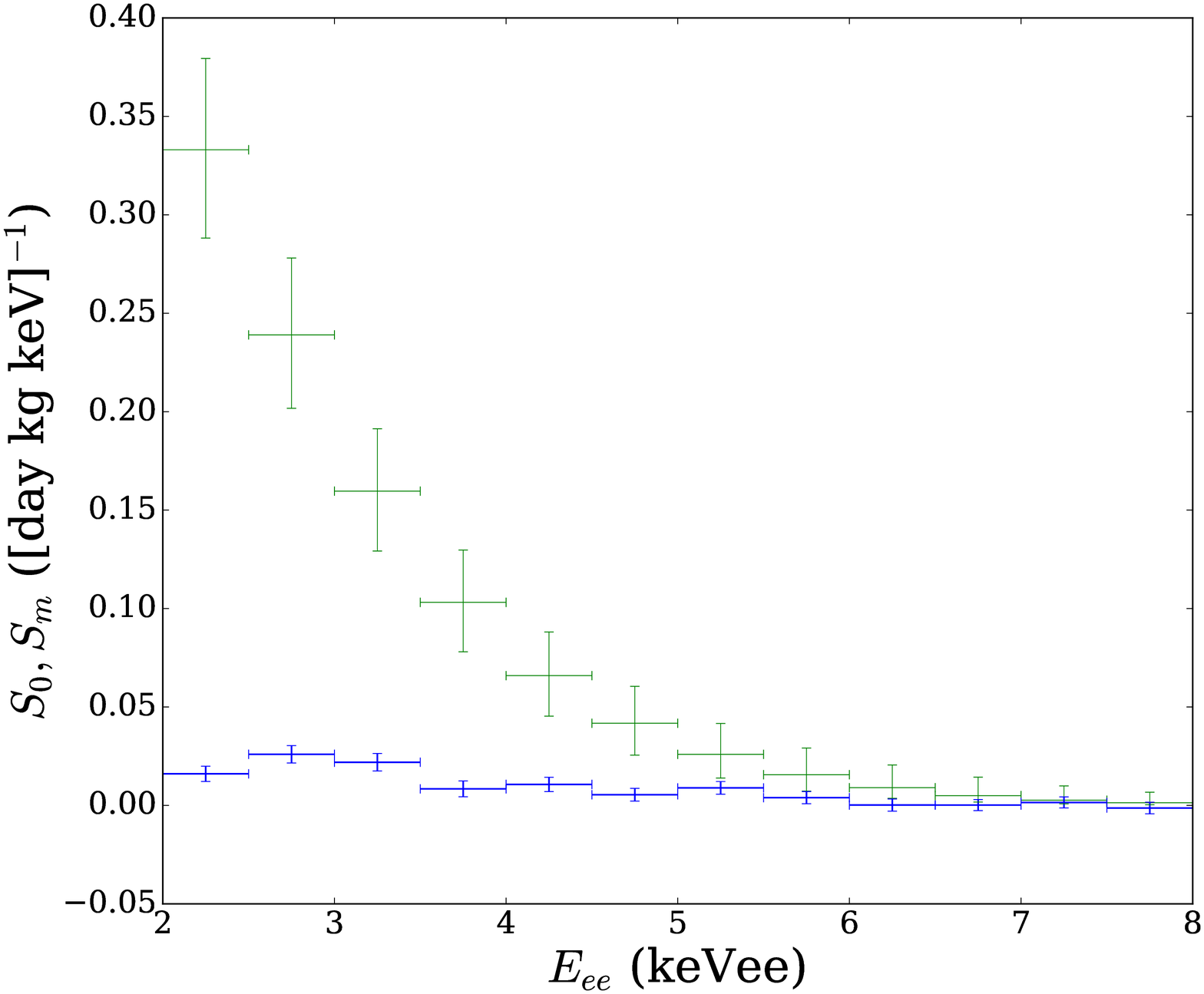}
\end{center}
\caption{Same as Fig.~\ref{fig:mchi_5_e_s0} but for $m_{\chi}$=10 GeV.}
\label{fig:mchi_10_e_s0}
\end{figure}

\begin{figure}[t]
\begin{center}
\includegraphics[width=0.8\columnwidth]{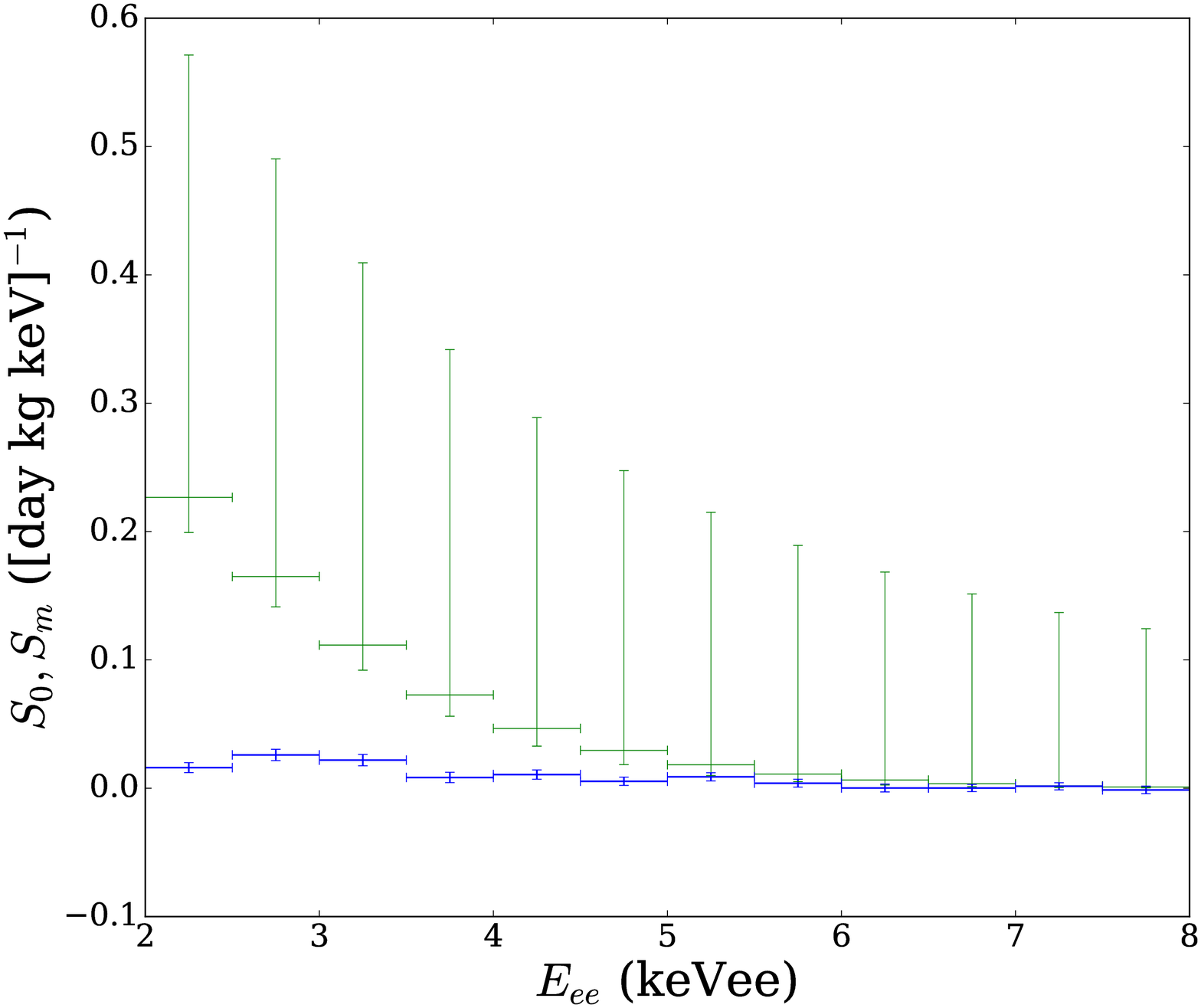}
\end{center}
\caption{Same as Fig.~\ref{fig:mchi_10_e_s0} but for $m_{\chi}$=15 GeV.}
\label{fig:mchi_15_e_s0}
\end{figure}

The maximum-likelihood estimates and the 1$\sigma$ confidence intervals of $S_{0,i}$ ($i=1,\ldots,12$), obtained from
Figs.~\ref{fig:mchi_5_profile},~\ref{fig:mchi_10_profile} and~\ref{fig:mchi_15_profile} (horizontal blue lines best visible in Fig.~\ref{fig:mchi_15_profile}),\footnote{For the higher masses, the sampling becomes sparse at large values of $S_{0,i}$ and the exact value of $S^{\rm sup}_{0,i}(L_0)$ becomes harder to determine from the figures. This problem is there because the MCMC concentrates points near the maximum-likelihood value of $S_{0,i}$. A few attempted methods of direct minimization of the $\chi^2$ at fixed $S_{0,i}$ did not converge. We have performed additional dedicated MCMC runs focused at large values of $S_{0,i}$ and the value of $S^{\rm sup}_{0,i}(L_0)$ did not change significantly.}
are listed in Table~\ref{tab:S}  and plotted versus the electron--equivalent
energy $E_{ee}$ in Figs.~\ref{fig:mchi_5_e_s0},~\ref{fig:mchi_10_e_s0}
and~\ref{fig:mchi_15_e_s0},  for $m_{\chi}=5$, 10, and 15 GeV, respectively. These are the main results of our paper.
Table~\ref{tab:S} also lists the DAMA modulated amplitudes $S_{m,i}$ (from Fig.~8 of~\cite{dama}) and the DAMA background+signal rates $B_i+S_i$ (from Fig.~27 of~\cite{DAMAbackground}, rebinned from 0.25-keVee- to 0.5-keVee-width bins). 

We see that for each $m_\chi$, the $S_{0,i}$ decrease with energy, as
reasonable for a WIMP signal. We also see that the error bars on the
$S_{0.i}$ are small enough to allow a rather good determination of the
unmodulated signal, although the error bars become very asymmetric for
$m_\chi = 15~{\rm GeV}$. Moreover, an examination of Table~\ref{tab:S}
leads to the conclusion that the unmodulated signals $S_{0,i}$ are
much smaller than the DAMA background+signal measurements. Since it is not trivial to
identify what contributes to the DAMA background, we refrain
from subtracting an estimated model background like done for instance
in~\cite{dama_bck}. It suffices for us to conclude that the $S_{0,i}$
values we estimate in this paper are reasonable and compatible with
the measured DAMA background+signal level.

\begin{table}[t]
\begin{center}
\def\arraystretch{1.5}
\begin{tabular}{|l|l|l|l|l|l|}
\hline
$E_i$&
$S_{m,i}$ &
$S_{0,i}$ &
$S_{0,i}$ &
$S_{0,i}$ &
$B_i+S_i$
\\[-1.5ex]
[keVee] &
&
$m_\chi=5$~GeV &
$m_\chi=10$~GeV &
$m_\chi=15$~GeV &
\\
\hline
 2.0--2.5  &  $0.0161(39)$  &  $0.513_{-0.050}^{+0.051}$  &  $0.333_{-0.045}^{+0.046}$  &  $0.227_{-0.027}^{+0.35}$  &  $1.029$  \\ 
 2.5--3.0  &  $0.0260(44)$  &  $0.311_{-0.030}^{+0.031}$  &  $0.239_{-0.037}^{+0.039}$  &  $0.165_{-0.024}^{+0.33}$  &  $1.228$  \\ 
 3.0--3.5  &  $0.0219(44)$  &  $0.165_{-0.016}^{+0.016}$  &  $0.160_{-0.030}^{+0.032}$  &  $0.112_{-0.020}^{+0.30}$  &  $1.294$  \\ 
 3.5--4.0  &  $0.0084(40)$  &  $0.0791_{-0.0078}^{+0.0078}$  &  $0.103_{-0.025}^{+0.027}$  &  $0.073_{-0.017}^{+0.27}$  &  $1.140$  \\ 
 4.0--4.5  &  $0.0107(36)$  &  $0.0352_{-0.0036}^{+0.0035}$  &  $0.066_{-0.021}^{+0.022}$  &  $0.047_{-0.014}^{+0.24}$  &  $0.956$  \\ 
 4.5--5.0  &  $0.0054(32)$  &  $0.0148_{-0.0016}^{+0.0015}$  &  $0.042_{-0.016}^{+0.019}$  &  $0.030_{-0.011}^{+0.22}$  &  $0.853$  \\ 
 5.0--5.5  &  $0.0089(32)$  &  $0.00600_{-0.0065}^{+0.0059}$  &  $0.026_{-0.012}^{+0.016}$  &  $0.018_{-0.0084}^{+0.20}$  &  $0.868$  \\ 
 5.5--6.0  &  $0.0039(31)$  &  $0.00236_{-0.00026}^{+0.00023}$  &  $0.0156_{-0.0084}^{+0.013}$  &  $0.011_{-0.0059}^{+0.18}$  &  $0.853$  \\ 
 6.0--6.5  &  $0.00018(308)$  &  $9.04_{-1.03}^{+0.89}\times10^{-4}$  &  $0.0090_{-0.0054}^{+0.011}$  &  $0.0064_{-0.0038}^{+0.16}$  &  $0.868$  \\ 
 6.5--7.0  &  $0.00018(281)$  &  $3.41_{-0.40}^{+0.34}\times10^{-4}$  &  $0.0050_{-0.0033}^{+0.0094}$  &  $0.0035_{-0.0023}^{+0.15}$  &  $0.860$  \\ 
 7.0--7.5  &  $0.0015(28)$  &  $1.27_{-0.15}^{+0.12}\times10^{-4}$  &  $0.0026_{-0.0018}^{+0.0073}$  &  $0.0019_{-0.0013}^{+0.13}$  &  $0.860$  \\ 
 7.5--8.0  & $-0.0013(29)$  &  $4.67_{-0.56}^{+0.46}\times10^{-5}$  &  $0.0013_{-0.0010}^{+0.0054}$  &  $0.95_{-0.71}^{+123}\times10^{-3}$  &  $0.890$  \\ 
\hline
\end{tabular}
\end{center}
\caption{DAMA modulation amplitudes $S_{m,i}$, estimated unmodulated rates $S_{0,i}$, and background+signal rates $B_i+S_i$ in counts/kg/day/keV.  Column 1: electron-equivalent energy bins. Column 2:
  DAMA modulation amplitudes $S_{m,i}$ from \cite{dama}. Columns 3--5:
  estimated unmodulated spectrum $S_{0,i}$ for WIMP masses
  $m_\chi=5~{\rm GeV}$, $m_\chi=10~{\rm GeV}$, and $m_\chi=15~{\rm
    GeV}$. Column 6: DAMA background+signal rates $B_i+S_i$ from
  \cite{DAMAbackground} (rebinned from 0.25-keVee- to 0.5-keVee-width
  bins).}
\label{tab:S}
\end{table}

\begin{figure}[t]
\begin{center}
\includegraphics[width=0.8\columnwidth]{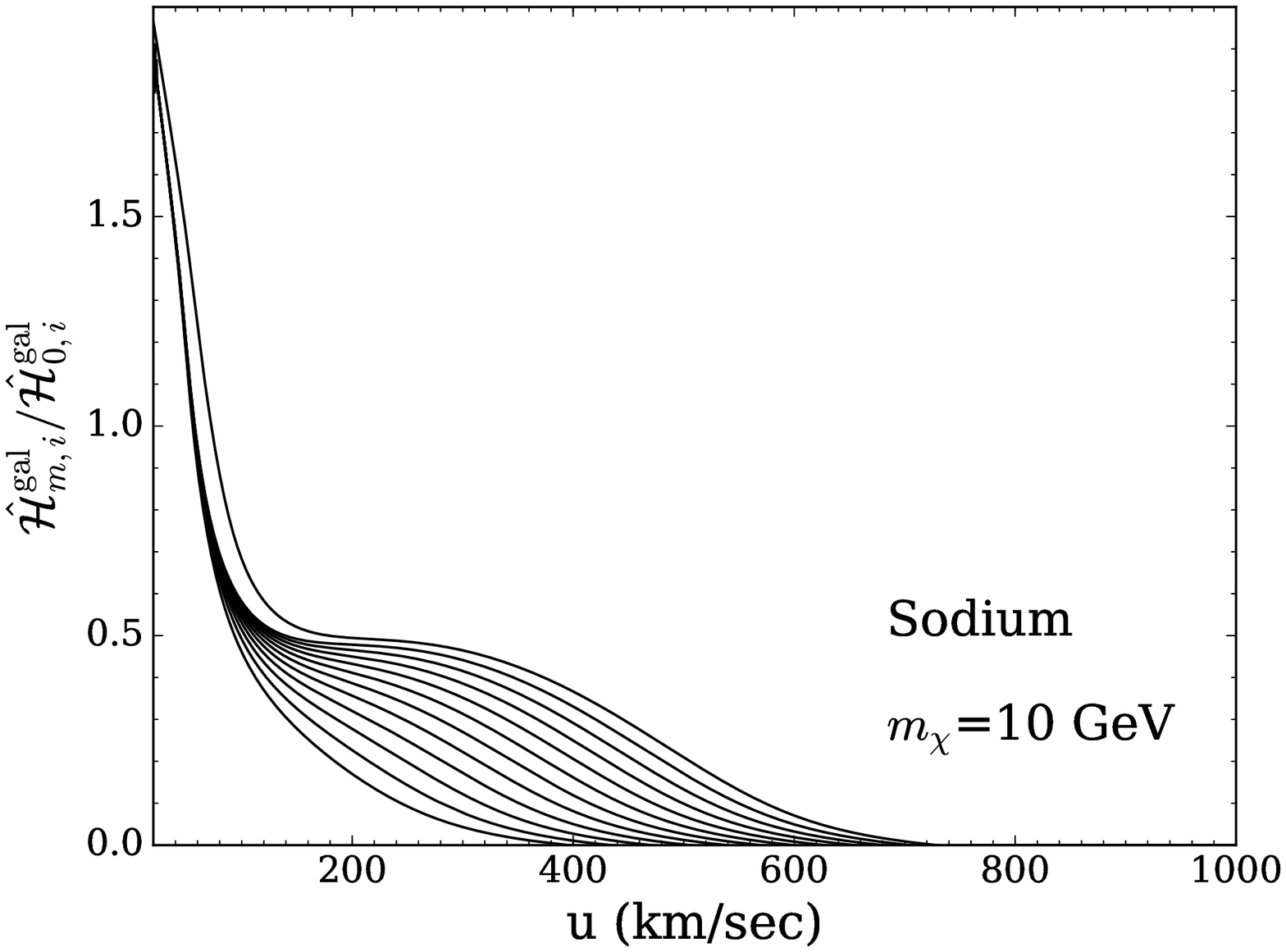}
\end{center}
\caption{Ratios $\scrH^{\gal}_{m,[E^{\prime}_1,E^{\prime}_2]}/\scrH^{\gal}_{0,[E^{\prime}_1,E^{\prime}_2]}$ of the modulated and
  unmodulated isotropic Galactic DAMA response functions shown in
  Figs.~\ref{fig:response_functions_hm_mchi_10} and
  \ref{fig:response_functions_h0_mchi_10} for $m_{\chi}$=10 GeV. Values larger than one
  signal the need of additional harmonics besides the cosine in the
  description of the modulation time dependence of the signal.
  \label{fig:response_functions_ratios_mchi_10}}
\end{figure}

It is finally interesting to estimate the fraction of the signal that
is modulated. In the energy range where the bulk of the DAMA
modulation is present, i.e., 2 keVee$<E_{ee}<$4 keVee, we find, for
$\chi^2_i\le\chi^2_{i,\rm min}+1$ (examining the ratio $S_{m,i}/S_{0,i}$ for each point in red in
Figs.~\ref{fig:mchi_5_profile}, \ref{fig:mchi_10_profile}
and~\ref{fig:mchi_15_profile}):
\begin{align}
& 0.04 \lsim \frac{S_{m,i}}{S_{0,i}} \lsim 0.14 \text{ for } m_\chi = 5~{\rm GeV}, 
\\
& 0.05 \lsim \frac{S_{m,i}}{S_{0,i}} \lsim 0.17 \text{ for } m_\chi = 10~{\rm GeV},
\\
& 0.03 \lsim \frac{S_{m,i}}{S_{0,i}} \lsim 0.24 \text{ for } m_\chi = 15~{\rm GeV}.
\end{align}
The modulated amplitude
ranges between a few percent and about 25\% of the unmodulated amplitude,
depending on the WIMP mass. This is in line with expectations for a signal due to dark matter WIMPs.

This conclusion is limited to the case at
study, i.e., to velocity distributions that are isotropic in the frame of the galaxy. However it is
worth pointing out that solutions with higher modulation fractions are
present in the allowed parameter space also in the isotropic case, but
are rejected by the DAMA data. This can be seen in
Fig.~\ref{fig:response_functions_ratios_mchi_10}, where we plot the ratios between
modulated and unmodulated response functions $\scrH^{\gal}_{m,i}/\scrH^{\gal}_{0,i}$ for $m_{\chi}$=10
GeV. In this specific example it is clear that, if $f_{\rm gal}(u)$ is
parametrized according to Eq.~(\ref{eq:fgalextreme}) with a set of $u_j$ velocities all below $\sim$60 km/s, modulation
fractions close to 1 are possible. These configurations, however, have a very small likelihood and do not appear in the 1$\sigma$ confidence intervals. Notice that at even lower
velocities, values of the $\scrH^{\gal}_{m,i}/\scrH^{\gal}_{0,i}$ ratios larger than one signal the need of
additional harmonics besides the cosine in the description of the
modulation time dependence of the signal.

\section{Conclusions}
\label{sec:conclusions}

We have estimated the unmodulated signal corresponding to the DAMA
modulation if interpreted as due to scattering of dark matter WIMPs
off ${}^{23}$Na in the DAMA detector. Our analysis, covering WIMPs
lighter than $\sim 15$ GeV, is to large extent independent of the dark
halo model (we profile the likelihood over velocity distributions that
are isotropic in the Galactic rest frame) and of the particle physics
model (we cover all ${}^{23}$Na--WIMP interactions in which the
${}^{23}$Na cross section vary little over the 2--4
keVee energy range where the DAMA modulation is significant).  The
method outlined in Section \ref{sec:extreme} is however very general
and valid for any class of velocity distribution.

We have presented and used an exact and sound mathematical set-up to
profile the likelihood over a continuum of nuisance parameters (the
whole WIMP velocity distribution). We have used the profile likelihood
for each of the unmodulated rates $S_{0,i}$ ($i=1,\ldots,12$ indexing
the first 12 DAMA energy bins) to find their maximum-likelihood
estimates and their 1$\sigma$ confidence intervals.

Our halo-independent estimates of the unmodulated rates are reasonable and in line with expectations for a signal from WIMP dark matter. The unmodulated rates we estimate give a modulated/unmodulated ratio ranging between a few percent and $\sim 25$\%. The unmodulated rates are comfortably below the background+signal level measured by DAMA.

\section*{Acknowledgements}

This work was completed during Stefano Scopel's sabbatical semester at
the University of Utah in the Spring of 2016, and was presented at
various conferences thereafter.  We thank the many audience members,
and in particular Riccardo Catena, for asking many questions that have
helped sharpening the writing of this paper. Paolo Gondolo thanks
Sogang University for the financial support and the kind hospitality
during his research visits. This work was partially supported by NSF
award PHY-1415974 at the University of Utah and by the National
Research Foundation of Korea grant number 2016R1D1A1A09917964 at
Sogang University.

\appendix

\section{Angle-averaged Galactic response functions}

In the application of the optimization method to the modulated/unmodulated DAMA rates in this paper we focus on velocity distributions that are isotropic in the Galactic frame, i.e., for which 
\begin{align}
f_\gal(\uu)=f_\gal(u),
\end{align}
where $u=|\uu|$. In this case, integrals of the form (\ref{eq:Si_of_t}) become
\begin{align}
S_{i}(t) = \int \overline{\scrH}^{\gal}_{i}(u,t) \, \overline{f}_\gal(u) \, du,
\end{align}
where
\begin{align}
\overline{\scrH}_{i}^{\gal}(u,t) = \frac{1}{4\pi} \int \scrH^{\gal}_{i}(\uu,t)  \, d\Omega_u
\label{eq:def_Hbargal}
\end{align}
is the angle-averaged Galactic response function, and 
\begin{align}
\overline{f}_{\gal}(u) = 4 \pi u^2 f_\gal(u) 
\end{align}
is the speed distribution in the Galactic frame normalized to one ($\int_0^\infty \overline{f}_{\gal}(u) \, du = 1$). A Fourier time-series analysis then gives, in this case of isotropic Galactic velocity distributions,
\begin{align}
& S_{0,i} = \int \overline{\scrH}^{\gal}_{0,i}(u) \, \overline{f}_\gal(u) \, du,
\\
& S_{m,i} = \int \overline{\scrH}^{\gal}_{m,i}(u) \, \overline{f}_\gal(u) \, du,
\end{align}
where
\begin{align}
& \overline{\scrH}^{\gal}_{0,i}(u) = \frac{1}{T} \int_{0}^{T} dt \, \overline{\scrH}^{\gal}_{i}(u,t) ,
\label{eq:def_Hbar0i}
\\
& \overline{\scrH}^{\gal}_{m,i}(u) = \frac{2}{T} \int_{0}^{T} dt \, \cos[\omega(t-t_0)] \, \overline{\scrH}^{\gal}_{i}(u,t) .
\label{eq:def_Hbarmi}
\end{align}

For nondirectional dark matter detectors, the laboratory response functions $\scrH_i(\vv)$ are isotropic, i.e., $\scrH_{i}(\vv) = \scrH_{i}(v)$. In this case, Eqs.~(\ref{eq:Si_of_t_gal_lab}), (\ref{eq:def_Hbargal}) and~(\ref{eq:def_Hbar0i}--\ref{eq:def_Hbarmi}) give
\begin{align}
\overline{\scrH}^{\gal}_{0,i}(u) & = \frac{1}{4\pi}  \int d\Omega_u \, \frac{1}{T} \int_0^{T} dt \, \scrH_i\big(|\uu-\VV|\big),
\\
\overline{\scrH}^{\gal}_{m,i}(u) & = \frac{1}{4\pi}  \int d\Omega_u \, \frac{2}{T} \int_0^{T} dt \, \cos(\omega t) \, \scrH_i\big(|\uu-\VV|\big),
\end{align}
where
\begin{align}
\VV & = \vvsun+\vvearth(t).
\end{align}
These integrals can be recast as integrals over $v=|\uu-\VV|$ with analytically computed kernels by means of the following manipulations. Let the polar angles $(\theta_u,\phi_u)$ of $\uu$ in $d\Omega_u=\sin\theta_u \, d\theta_u \, d\phi_u$ be defined with respect to the direction of $\VV$. Then 
\begin{align}
v & = |\uu-\VV| = \sqrt{u^2+V^2-2uV\cos\theta_u} .
\end{align}
Perform the trivial integration over $\phi_u$, and change integration variable from $\cos\theta_u$ to $v$. Then interchange the order of the $t$ and $v$ integrations. This gives
\begin{align}
\overline{\scrH}^{\gal}_{0,i}(u) = \int_{0}^{\infty} dv \, v \, \scrH_i(v) \, \frac{1}{T} \int_{0}^{T} dt \, \frac{1}{2uV}
\, \Theta\big( |u-v| \le V \le u+v \big) ,
\label{eq:H0caux2}
\\
\overline{\scrH}^{\gal}_{m,i}(u) = \int_{0}^{\infty} dv \, v \, \scrH_i(v) \, \frac{2}{T} \int_{0}^{T} dt \, \frac{\cos(\omega t)}{2uV}
\, \Theta\big( |u-v| \le V \le u+v \big) .
\label{eq:Hmcaux2}
\end{align}
Here $\Theta(x)=1$ if $x$ is true and $\Theta(x)=0$ if $x$ is false. The integrals over the time $t$ in Eqs.~(\ref{eq:H0caux2}) and~(\ref{eq:Hmcaux2}) can be performed analytically for a circular Earth orbit. Write 
\begin{align}
\vvearth(t) = \vv_{\earth 1} \cos(\omega t) + \vv_{\earth 2} \sin(\omega t),
\end{align}
where $t=0$ when $\vvearth$ and $\vvsun$ form their smallest angle $\beta$, i.e., $\vvsun\cdot\vv_{\earth 1} = \vsun\vearth\cos\beta$ and $\vvsun\cdot\vv_{\earth 2} = 0$. Then $\vvsun\cdot\vv_{\earth}(t) = \vsun\vearth\cos\beta\cos(\omega t)$ and
\begin{align}
V = \sqrt{\vsun^2+\vearth^2+2\vsun\vearth\cos\beta\cos(\omega t)} = v_a \sqrt{\strut 1+\epsilon \cos(\omega t)},
\end{align}
where 
\begin{align}
v_a = \sqrt{\vsun^2+\vearth^2},
\qquad
\epsilon = \frac{2\vsun\vearth\cos\beta}{\vsun^2+\vearth^2}.
\end{align}
In terms of these variables,
\begin{align}
\overline{\scrH}^{\gal}_{0,i}(u) & = \frac{1}{2uv_a} \int_{0}^{\infty} dv \, v \, \scrH_i(v) \, K_0\bigg( \frac{u}{v_a}, \frac{v}{v_a} \bigg) \label{eq:h0_fourier},
\\
\overline{\scrH}^{\gal}_{m,i}(u) & = - \frac{\vsun\vearth\cos\beta}{2uv_a^3} \int_{0}^{\infty} dv \, v \, \scrH_i(v) \, K_m\bigg( \frac{u}{v_a}, \frac{v}{v_a} \bigg) \label{eq:hm_fourier},
\end{align}
where we define
\begin{align}
K_0(x,y) & = \frac{1}{T} \int_0^{T} dt \, \frac{1}{\sqrt{1+\epsilon\cos(\omega t)}} \Theta\big( (x-y)^2 \le 1+ \epsilon\cos(\omega t) \le (x+y)^2 \big) ,
\\
K_m(x,y) & = - \frac{4}{\epsilon \, T} \int_0^{T} dt \, \frac{\cos(\omega t)}{\sqrt{1+\epsilon\cos(\omega t)}} \Theta\big( (x-y)^2 \le 1+ \epsilon\cos(\omega t) \le (x+y)^2 \big) .
\end{align}
The functions $K_0(x,y)$ and $K_m(x,y)$ are symmetric in $x$ and $y$, and the integrals appearing in them can be expressed in terms of the incomplete elliptic functions $E(\phi|m)$ and $F(\phi|m)$. 
\begin{align}
K_0(x,y) & = I_0\big(\alpha(x-y)\big)-I_0\big(\alpha(x+y)\big),
\\
K_m(x,y) & = I_m\big(\alpha(x-y)\big)-I_m\big(\alpha(x+y)\big),
\end{align}
where
\begin{align}
I_0(\alpha) & = \frac{1}{\pi \sqrt{1+\epsilon}} \, F\bigg(\frac{\alpha}{2}\bigg|\frac{2\epsilon}{1+\epsilon}\bigg) ,
\\
I_m(\alpha) & = - \frac{8}{\pi \epsilon\sqrt{1+\epsilon}} \left[ (1+\epsilon) E\bigg(\frac{\alpha}{2}\bigg|\frac{2\epsilon}{1+\epsilon}\bigg) - F\bigg(\frac{\alpha}{2}\bigg|\frac{2\epsilon}{1+\epsilon}\bigg) \right],
\end{align}
and
\begin{align}
\alpha(\xi) = 
\begin{cases}
\pi, & |\xi|\le\sqrt{1-\epsilon} ,\\
\arccos\bigg(\frac{\xi^2-1}{\epsilon}\bigg), & \sqrt{1-\epsilon} \le |\xi| \le \sqrt{1+\epsilon}, \\
0, & |\xi|\ge\sqrt{1+\epsilon} .
\end{cases}
\end{align}
We used Eqs.~(\ref{eq:h0_fourier}) and~(\ref{eq:hm_fourier}) to compute the DAMA angle-averaged Galactic response functions.

We notice lastly that the angle--averaged Galactic response functions for the $\sin(\omega t)$ Fourier mode vanish identically. This follows from the observation that the formula for the angle--averaged sine modulation response functions is analogous to equation (\ref{eq:Hmcaux2}) with the replacement of $\cos(\omega t)$ with $\sin(\omega t)$ in the numerator but the same dependence on $V$. Since $V$ depends on $\cos(\omega t)$ only, the integrand is thus a product of an odd and an even function of $\omega t$ and vanishes when integrated over a whole period.


\end{document}